\documentclass[aps,preprint]{revtex4}%
\usepackage{amsfonts}
\usepackage{amsmath}
\usepackage{amssymb}
\usepackage{graphicx}%
\providecommand{\U}[1]{\protect\rule{.1in}{.1in}}
\newtheorem{theorem}{Theorem}

\newtheorem{conjecture}[theorem]{Conjecture}

\begin{document}
\preprint{NENP-C}
\title[ ]{A Possible Universal Model without Singularity and its Explanation for
Evolution of the Universe}
\author{Shi-Hao Chen}
\affiliation{Institute of Theoretical Physics, Northeast Normal University, Changchun
130024, China.}

\begin{abstract}
The new hypotheses are proposed that there are $s-particles$ and $v-particles$
which are symmetric and mutually repulsive, there are $S-space$ and $V-space$
whose essential difference is only that their expectation values of the Higgs
fields are different. In $S-space$ the $S-SU(5)$ symmetry is broken into
$S-SU(3)XU\left(  1\right)  ,$ and $V-SU(5)$ symmetry still holds$.$ As a
consequence, $s-particles$ get their masses determined by the $SU(5)$ $GUT$
and form the $S-world$, and $v-particles$ are all massless and form $SU(5)$
colour-single states which are identified with dark energy$.$ The following
results are obtained. There is no spacetime singularty, and there is the
highest temperature in the universe. The creating process of one world is just
the annihilating process of the other world in the highest temperature. A
formula which well describes the luminous distance and redshift is
obtained.\ The results of the Guth's inflationary scenario are obtained.
Decelerated expanding early stage and accelerated expanding now stage of the
universe are explained. New predictions are follows. Some huge cavities in
$V-space$ are not empty, in which there is $s-matter$ with larger density, and
are equivalent to huge concave lenses. The given tharacters of some huge
cavities are well explained. The gravitation between two galaxies distant
enough will lesser than that predicted by the conventional theory. A possible
explanation for the big redshift of quasi-stellar objects is presented. Huge
redshifts of quasars are mass redshifts. The universe is composed of infinite
universal islands. It is possible that there is a new annihilating mode of
black holes with their very huge masses, and there are very huge white holes
which are different from that predicted by conventional theory.

$\mathbf{PACS}$\textbf{: }$04.50.+h$, $98.80.Cq$.


\end{abstract}
\volumeyear{year}
\volumenumber{number}
\issuenumber{number}
\eid{identifier}
\date[Date text]{date}
\received[Received text]{date}

\revised[Revised text]{date}

\accepted[Accepted text]{date}

\published[Published text]{date}

\startpage{1}
\endpage{ }
\maketitle
\tableofcontents

\section{Introduction}

As is now well known, there is space-time singularity under certain
conditions$^{[1]}$. These conditions fall into three categories. First, there
is the requirement that gravity shall be attractive. Secondly, there is the
requirement that there is enough matter present in some region to prevent
anything escaping from that region. The third requirement is that there should
be causality violations. The recent astronomical observations show that the
universe expanded with a deceleration early and is expanding with an
acceleration now. This implies that there is dark energy. $0.73$ of the total
energy of the universe is dark energy$^{[2]}.$ What is dark energy? Many
possible answers have been given. One possible interpretation is in terms of
the effective cosmological constant $\lambda_{eff}=$ $\lambda+E_{0},$ here
$\lambda$ and $E_{0}$ are respectively the Einstein's cosmological constant
and the zero-point energy of the vacuum. But $\lambda_{eff}$ cannot been
derived from basic theory $[3]$ and $E_{0}\gg\lambda_{eff}$. Hence the
interpretation is unsatisfactory. Alternatively, dark energy is associated
with the dynamics of scalar field $\phi\left(  t\right)  $ that is uniform in
space$^{[4]}.$This is a seesaw cosmology$^{[5]}.$ This implies that the
conventional theory is still unsatisfied and the issue of the cosmological
constant is still unsolved.

We hope to solve the two problems above on the premise $E_{0}=\lambda=0.$ We
suppose $\lambda_{eff}=0.$ On the premise and the basis of a new conjecture we
try to solve the problems of space-time singularity and the universe expanding
with an acceleration. In fact, a sort of quantum field theory has been
constructed in which there is no divergence, and $E_{0}=0^{[6]}.$ But the
present model is not on the basis of the quantum field theory without
divergence. Hence we suppose $\lambda_{eff}=0.$ The other premise of the
present model is the conventional $SU(5)$ grand unified theory $\left(
GUT\right)  $ But it is easily seen that the present model does not rely the
especial $GUT$ model. In fact, provided there are the $GUT$ model, and when
the symmetry strictly hold water, there are the glue-balls corresponding to
symmetry, the present model can hold water. As is well known, although the
glue-balls corresponding to $SU(3)$ are not founded for a time, the $SU(3)$
theory have proven that there can be the $SU(3)$ glue-balls whose masses are
not zero.

According to the present model there are two sorts of matter which are called
$s-matter$ and $v-matter$ and which are repulsive and symmetric each other,
hence one condition of the space-time singularity theorems is violated.
Consequently, there is no the space-time singularity. One of both sorts of
matter ($s-matter$) can be identified with the conventional matter forming the
given world, and other sort ($v-matter$) can be identified with the so-called
dark energy. The results of the Guth's inflationary scenario are obtained.
That the universe expanded with a deceleration early and is expanding with an
acceleration now is explained are explained. The present model have the
following new predictions. There is the highest temperature in the universe,
hence there is no space-time singularity. That $S-world$ come into being
follows that $V-world$ disappears at the highest temperature. There are the
repulsion red-shifts and the mass red-shifts which are essentially different
from the given red-shifts. Some $s-huge$ cavities in $S-world$ are not empty,
but there is $v-matter$ in them, their effects are similar to that of huge
concave lens. Because there must be $v-matter$ between two $s-galaxies$ when
the distant between both is large enough, the gravitation between both will
lesser than that predicted by the conventional theory. According to the model,
the universe is composed of infinite universal islands. From this we give a
possible explanation to the very large red-shift of a quasar. The huge
redshifts of quasars are the mass redshifts. It is possible that a $v-black$
hole with its mass big enough can transform into $s-huge$ white hole which
cannot be observed by a $v-observer$ except by its repulsion effects, and the
$s-cosmic$ island neighboring on a $v-cosmic$ island transforms into a
$v-cosmic$ island after it contraction. In this case, a $v-observer$ in the
$v-cosmic$ island must observe a very huge $v-huge$ white holes which are
different from that predicted by conventional theory.

Before inflating occurs, the $s-world$ is in thermal equilibrating state. If
there is no $v-matter,$ because of contraction by gravitation, the $s-world$
would become a thermal equilibrating singular point, i.e., the $s-world$ would
be in the hot death state. As seen, it is necessary that there are both
$s-matter$ and $v-matter$ and both $S-space$ and $V-space$.

\section{Action, energy-momentum tensor and equations of motion}

In order to solve the two problems of singular point and dark energy, we
propose the following conjectures.

\begin{conjecture}
: There are two sorts of matter which are respectively called $solid-matter$
($s-matter$) and $void-matter$ ($v-matter$). $s-matter$ and $v-matter$ are
symmetric; Contribution of $s-matter$ to the Einstein tensor is opposite to
that of $v-matter;$ There is no other interaction between both except there is
the interaction described by $(13)$ between $s-Higgs$ $fields$ and $v-Higgs$
$fields$.
\end{conjecture}

Matter determines properties of spacetime. Different breaking modes of Higgs
fields correspond to different ground states. The Higgs fields in the present
model have two sorts of breaking modes different in essence. One sort is that
the vacuum expectations of the $s-Higgs$ $fields$ are not zero and all the
vacuum expectations of the $v-Higgs$ $fields$ are zero. In the case, space is
called $Solid-space$ ($S-space$). The other sort is that all the vacuum
expectations of the $s-Higgs$ $fields$ are zero and the vacuum expectations of
the $v-Higgs$ $fields$ are not zero. In the case, space is called $Void-space$
($V-space$). We will see that in the $S-space$ $s-particles$ can get their
masses and the masses of all $v-particles$ are zero (hence, $v-particles$ are
identified with dark energy in $S-space$) ,\ in the $V-space$ $v-particles$
can get their masses and the masses of all $s-particles$ are zero (hence,
$s-particles$ are identified with dark energy in the $V-space$), and one sort
of space can transform into the other as changing of breaking modes of Higgs fields.

Because contribution of $s-matter$ to the Einstein tensor is opposite to that
of $v-matter$ and there is no other interaction, there is only repulsion
between $s-matter$ and $v-matter$. Thus, it is obvious that $s-objects$
($v-objects$) is composed of only $s-particles$ (only $v-particles$). If both
sorts of matter exist in the same form (e.g., both exist as atoms) and
simultaneously exist in the same region, the hypothesis will be in
contradiction with given observations. But this is impossible because one of
both must be identified with dark energy which cannot be observed and the
other must be identified with matter in the given world. Thus the hypothesis
is not in contradiction with all given experiments and astronomical observations.

Of course, in fact, only there is one sort of space and one sort of metric
tensors $g_{\mu\nu}$ describing the structure of the space. In order to
emphasize the important significance of the vacuum expectations of the Higgs
fields, we define $S-space$ and $V-space.$ The essential difference of
$S-space$ and $V-space$ is that the vacuum expectations of the Higgs fields in
$S-space$ are different from those in $V-space.$ Consequently the existing
forms of matter and the metric tensors in $S-space$ must be different from
those in $V-space.$

Because the interaction between the $s-matter$ and the $v-matter$ is repulsive
each other, one condition of the space-time singularity theorems is violated.
Consequently, there is no the space-time singularity according to the present model.

We denote\ physical quantities in $S-space$ (in $V-space)$ by subscript `$S$'
( `$V$' ), and denote physical quantities determined by $s-matter$
($v-matter)$ by subscript `$s$' ( `$v$' ). But it is possible that the
subscripts $S$ and $V$ are not marked, if there is no confusion.

On the basis of the conjecture above the actions should be written as two
forms, $I_{S}$ in the $S-space$ and $I_{V}$ in the $V-space,$ respectively, at
the 0-temperature.%
\begin{equation}
I_{S}=\int d^{4}x_{S}\sqrt{-g_{S}}\mathcal{L}_{S}=\int d^{4}x_{V}\sqrt{-g_{V}%
}\mathcal{L}_{V}=I_{V}, \tag{1}%
\end{equation}%
\begin{equation}
\mathcal{L}_{S}=\frac{1}{16\pi G}R_{S}+s\mathcal{L}_{Ss}+v\mathcal{L}%
_{Sv}+\frac{1}{2}\left(  v+s\right)  V_{Ssv}, \tag{2}%
\end{equation}%
\begin{equation}
\mathcal{L}_{Ss}=\mathcal{L}_{sm}\left(  \Psi_{s}\left(  x_{S}\right)
,g_{S}\left(  x_{S}\right)  ,g_{S}\left(  x_{S}\right)  ,_{\mu}\right)
+V_{s}\left(  \omega_{s}\left(  x_{S}\right)  \right)  +V_{0}, \tag{3}%
\end{equation}%
\begin{equation}
\mathcal{L}_{Sv}=\mathcal{L}_{vm}\left(  \Psi_{v}\left(  x_{S}\right)
,g_{S}\left(  x_{S}\right)  ,g_{S}\left(  x_{S}\right)  ,_{\mu}\right)
+V_{v}\left(  \omega_{v}\left(  x_{S}\right)  \right)  , \tag{4}%
\end{equation}%
\begin{equation}
V_{Ssv}=V_{sv}\left(  \omega_{s}\left(  x_{S}\right)  ,\omega_{v}\left(
x_{S}\right)  \right)  ; \tag{5}%
\end{equation}%
\begin{equation}
\mathcal{L}_{V}=\frac{1}{16\pi G}R_{V}+v\mathcal{L}_{Vs}+s\mathcal{L}%
_{Vv}+\frac{1}{2}\left(  v+s\right)  V_{Vsv}, \tag{6}%
\end{equation}%
\begin{equation}
\mathcal{L}_{Vs}=\mathcal{L}_{sm}\left(  \Psi_{s}\left(  x_{V}\right)
,g_{V}\left(  x_{V}\right)  ,g_{V}\left(  x_{V}\right)  ,_{\mu}\right)
+V_{s}\left(  \omega_{s}\left(  x_{V}\right)  \right)  , \tag{7}%
\end{equation}%
\begin{equation}
\mathcal{L}_{Vv}=\mathcal{L}_{vm}\left(  \Psi_{v}\left(  x_{V}\right)
,g_{V}\left(  x_{V}\right)  ,g_{V}\left(  x_{V}\right)  ,_{\mu}\right)
+V_{v}\left(  \omega_{v}\left(  x_{V}\right)  \right)  +V_{0}, \tag{8}%
\end{equation}%
\begin{equation}
V_{Vsv}=V_{sv}\left(  \omega_{s}\left(  x_{V}\right)  ,\omega_{v}\left(
x_{V}\right)  \right)  , \tag{9}%
\end{equation}%
\begin{equation}
\omega_{s}\equiv\Omega_{s},\;\Phi_{s},\;\chi_{s};\;\ \ \ \ \ \omega_{v}%
\equiv\Omega_{v},\;\Phi_{v},\;\chi_{v}, \tag{10}%
\end{equation}
where the meanings of the symbols are follows. $g=\det(g_{\mu\nu})$,
$g_{\mu\nu}=diag(-1,1,1,1)$ in flat space. $g=g_{S}$ or $g_{V}$. $R_{S}$ and
$R_{V}$ are respectively the scalar curvatures in $S-space$ and $V-space$. $s$
and $v$ are two parameters and we finally take $s=-v=1$. $V_{0}$ is a
parameter which is so taken that $\left(  V_{s}\left(  \omega_{s}\left(
x_{S}\right)  \right)  +V_{0}\right)  _{\min}=\left(  V_{v}\left(  \omega
_{v}\left(  x_{V}\right)  \right)  +V_{0}\right)  _{\min}=0$ at the
0-temperature. $\mathcal{L}_{sm}$ ($\mathcal{L}_{vm}$) is the Lagrangian
density of all $s-fields$ ($v-fields$) and their couplings of the $SU(5)$
$GUT$ except the Higgs potentials $V_{s},$ $V_{v}$ and $V_{sv}$. $\Psi_{s}$
and $\Psi_{v}$ represents all $s-fields$ and all $v-fields$, respectively.
$\mathcal{L}_{s}$ and $\mathcal{L}_{s}$ do not contain the contribution of the
gravitational and repulsive fields. It is seen from $(2)-(9)$ that provided
the\quad subscripts\quad$S$\quad and\quad$V$\ in\quad$(2)-(5)$ are
interchanged, $(6)-(9)$ are obtained. $\mathcal{L}_{Ss}$ and $\mathcal{L}%
_{Vs}$ are different denoting modes of the same Lagrangian density in
different spaces.
\begin{align}
V_{s}  &  =-\frac{1}{2}\mu^{2}\Omega_{s}^{2}+\frac{1}{4}\lambda\Omega_{s}%
^{4}\nonumber\\
&  -\frac{1}{2}\eta\Omega_{s}^{2}Tr\Phi_{s}^{2}+\frac{1}{4}a\left(  Tr\Phi
_{s}^{2}\right)  ^{2}+\frac{1}{2}bTr\left(  \Phi_{s}^{4}\right) \nonumber\\
&  -\frac{1}{2}\varsigma\Omega_{s}^{2}\chi_{s}^{+}\chi_{s}+\frac{1}{4}%
\xi\left(  \chi_{s}^{+}\chi_{s}\right)  ^{2}, \tag{11}%
\end{align}

\begin{align}
V_{v}  &  =-\frac{1}{2}\mu^{2}\Omega_{v}^{2}+\frac{1}{4}\lambda\Omega_{v}%
^{4}\nonumber\\
&  -\frac{1}{2}\eta\Omega_{v}^{2}Tr\Phi_{v}^{2}+\frac{1}{4}a\left(  Tr\Phi
_{v}^{2}\right)  ^{2}+\frac{1}{2}bTr\left(  \Phi_{v}^{4}\right) \nonumber\\
&  -\frac{1}{2}\varsigma\Omega_{v}^{2}\chi_{v}^{+}\chi_{v}+\frac{1}{4}%
\xi\left(  \chi_{v}^{+}\chi_{v}\right)  ^{2}, \tag{12}%
\end{align}%
\begin{align}
V_{sv}  &  =\frac{1}{2}\Lambda\Omega_{s}^{2}\Omega_{v}^{2}+\frac{1}{2}%
\alpha\Omega_{s}^{2}Tr\Phi_{v}^{2}+\frac{1}{2}\beta\Omega_{s}^{2}\chi_{v}%
^{+}\chi_{v}\nonumber\\
&  +\frac{1}{2}\alpha\Omega_{v}^{2}Tr\Phi_{s}^{2}+\frac{1}{2}\beta\Omega
_{v}^{2}\chi_{s}^{+}\chi_{s}, \tag{13}%
\end{align}
where $\Omega_{s}$ and $\Omega_{v}$, $\Phi_{s}=\overset{24}{\underset
{i=1}{\sum}}\frac{T_{i}}{\sqrt{2}}\varphi_{i}$ and $\Phi_{v},$ $\chi_{s}$ and
$\chi_{v}$ are respectively $\underline{1}$ , $\underline{24}$\ and
$\underline{5}$ dimensional representations of the $SU(5)$ group, here the
couplings of $\Phi_{s}$ and $\chi_{s}$ ($\Phi_{v}$ and $\chi_{v}$) are ignored
for short$^{[7]}$ $.$

We will see that the present model takes $SU(5)$ $GUT$ \ only as an example.
In fact, the consequences of the present model is not dependent on specific
model, provided a model contains such Higgs couplings as $(11)-(13)$.

From $(1)$ the energy-momentum tensor densities which do not contain the
energy-momentum tensor of gravitational and repulsive fields can be defined
as\bigskip%
\begin{align}
T_{S\mu\nu}  &  =\frac{2}{\sqrt{-g_{S}}}\left(  \frac{\partial}{\partial
s}+\frac{\partial}{\partial v}\right)  \left[  \frac{\partial\left(
\sqrt{-g_{S}}\mathcal{L}_{S}\right)  }{\partial g_{S}^{\mu\nu}}-\left(
\frac{\partial\left(  \sqrt{-g_{S}}\mathcal{L}_{S}\right)  }{\partial
g_{S}^{\mu\nu},_{\sigma}}\right)  ,_{\sigma}\right] \nonumber\\
&  \equiv T_{Ss\mu\nu}+T_{Sv\mu\nu}+T_{Ssv\mu\nu}, \tag{14}%
\end{align}%
\begin{align}
T_{Ss\mu\nu}  &  =\,T_{Ssm\mu\nu}+\,T_{Ssv\mu\nu}=\frac{2}{\sqrt{-g_{S}}%
}\left[  \frac{\partial\left(  \sqrt{-g_{S}}\mathcal{L}_{sm}\right)
}{\partial g_{S}^{\mu\nu}}-\left(  \frac{\partial\left(  \sqrt{-g_{S}%
}\mathcal{L}_{sm}\right)  }{\partial g_{S}^{\mu\nu},_{\sigma}}\right)
,_{\sigma}\right] \nonumber\\
&  \text{ \ \ }-g_{S\mu\nu}\left(  V_{s}+V_{0}\right)  , \tag{15}%
\end{align}%
\begin{equation}
T_{Sv\mu\nu}=T_{Svm\mu\nu}-\,T_{Svv\mu\nu}=\frac{2}{\sqrt{-g_{S}}}\left[
\frac{\partial\left(  \sqrt{-g_{S}}\mathcal{L}_{vm}\right)  }{\partial
g_{S}^{\mu\nu}}-\left(  \frac{\partial\left(  \sqrt{-g_{S}}\mathcal{L}%
_{vm}\right)  }{\partial g_{S}^{\mu\nu},_{\sigma}}\right)  ,_{\sigma}\right]
-g_{S\mu\nu}V_{v}, \tag{16}%
\end{equation}%
\begin{equation}
T_{Ssv\mu\nu}=-g_{S\mu\nu}V_{sv} \tag{17}%
\end{equation}

\begin{align}
&  Interchanging\quad the\quad subscripts\quad S\quad and\quad
V\;and\;the\quad subscripts\quad s\quad and\quad v\;\nonumber\\
&  in\quad(14)-(17),we\quad get\quad\;T_{V\mu\nu},\,\;T_{Vsm\mu\nu
},\,\;T_{Vvm\mu\nu},\;T_{Vsv\mu\nu},\;\,T_{Vvv\mu\nu},\,\;T_{VS\mu\nu}.
\tag{18}%
\end{align}

Einstein's equation of gravitation field can be derived from $(1)$ \ \ \ \ \ \ \ \ \ \ \ \ %

\begin{equation}
R_{S\mu\nu}-\frac{1}{2}g_{S\mu\nu}R_{S}=-8\pi G\left(  T_{Ss\mu\nu}%
-T_{Sv\mu\nu}\right)  , \tag{19}%
\end{equation}

\begin{equation}
R_{V\mu\nu}-\frac{1}{2}g_{V\mu\nu}R_{V}=-8\pi G\left(  T_{Vv\mu\nu}%
-T_{Vs\mu\nu}\right)  . \tag{20}%
\end{equation}
It must be emphasized that\ the equations $(19)+(20)$ or $T_{Ss\mu\nu
}+T_{Vv\mu\nu}$ are all nonsensical, since $R_{S\mu\nu}$ and $R_{V\mu\nu}$ are
respectively the curvature tensors in the two different spaces, $T_{Ss\mu\nu}$
and $T_{Vv\mu\nu}$ are physical quantities in the two different spaces,
respectively, and both $S-space$ and $V-space$ cannot simultaneously exist.
Only one of the equations $(19)$ and $(20)$ corresponds to real space and is
applicable. Similarly, only one of $I_{S\text{ \ }}$and $I_{V\text{ \ }}$is
applicable. It should be noticed from $(19)-(20)$ and $(14)-(18)$ that the
potential energy $V_{sv}$ is different from other energies in essence,
$V_{sv}$ does not influence $R_{S\mu\nu}$ and $R_{V\mu\nu},$ i.e., there is no
gravitation and repulsion of the potential energy $V_{sv}.$ But this does not
cause any contradiction with all given experiments and astronomical
observations since $V_{sv}=0$ when $\langle\omega_{s}\rangle_{0}=0$ in the
$V-space$ or $\langle\omega_{v}\rangle_{0}=0$ in the $S-space$.

It is proven that the necessary and sufficient condition of $T_{;\nu}^{\mu\nu
}=0$ is $I_{M}$ to be a scalar quantity$^{[8]}$. Form $\left(  1\right)
-\left(  18\right)  $ we see that $I_{S}$ and $I_{V}$ are all scalar
quantities, hence when the energy-momentum tensors of the gravitational field
and repulsive field are not considered,
\[
\left(  T_{Ss}^{\mu\nu}+T_{Sv}^{\mu\nu}\right)  _{;\nu}=\left(  T_{Vs}^{\mu
\nu}+T_{Vv}^{\mu\nu}\right)  _{;\nu}=0.
\]

Because there is repulsive interaction between both $s-matter$ and $v-matter$,
the motional law of $v-matter$ must be different from the motional law of
$s-matter$ in the same gravitational fields. According to the general
relativity, it is obtained from $(19)$ the motional equation of $s-matter$ in
$S-space$ to be
\begin{equation}
\frac{d^{2}x_{S}^{\mu}}{d\sigma^{2}}+\Gamma_{S\alpha\beta}^{\mu}\frac
{dx_{S}^{\alpha}}{d\sigma}\frac{dx_{S}^{\beta}}{d\sigma}=0, \tag{21}%
\end{equation}
where $\sigma$ is a scalar parameter satisfying a proper equation. In contrast
with $s-matter$, the motional equation of $v-matter$ in $S-space$ will is
\begin{equation}
-\frac{d^{2}x_{S}^{\mu}}{d\sigma^{2}}+\Gamma_{S\alpha\beta}^{\mu}\frac
{dx_{S}^{\alpha}}{d\sigma}\frac{dx_{S}^{\beta}}{d\sigma}=0. \tag{22}%
\end{equation}
Analogously, we obtain the motional equation of $v-matter$ in $V-space$ to be%
\begin{equation}
\frac{d^{2}x_{V}^{\mu}}{d\sigma^{2}}+\Gamma_{V\alpha\beta}^{\mu}\frac
{dx_{V}^{\alpha}}{d\sigma}\frac{dx_{V}^{\beta}}{d\sigma}=0, \tag{23}%
\end{equation}
and the motional equation of $s-matter$ in $V-space$ to be
\begin{equation}
-\frac{d^{2}x_{V}^{\mu}}{d\sigma^{2}}+\Gamma_{V\alpha\beta}^{\mu}\frac
{dx_{V}^{\alpha}}{d\sigma}\frac{dx_{V}^{\beta}}{d\sigma}=0. \tag{24}%
\end{equation}

\section{Spontaneous breaking of symmetry}

Ignoring the couplings of $\Phi_{s}$ and $\chi_{s\text{ }}$and suitably
choosing the parameters of the Higgs potential$,$ analogously to $[7]$ and
$[9]$ we can prove from $(11)-(13)$ that there are the following vacuum
expectation values,%

\begin{equation}
\left\langle 0\left\vert \Omega_{v}\right\vert 0\right\rangle =\overline
{\Omega}_{v0}=\left\langle 0\left\vert \Phi_{v}\right\vert 0\right\rangle
=\overline{\Phi}_{v0}=\left\langle 0\left\vert \chi_{v}\right\vert
0\right\rangle =\overline{\chi}_{v0}=0, \tag{25}%
\end{equation}

\begin{equation}
\left\langle 0\left\vert \Omega_{s}\right\vert 0\right\rangle =\overline
{\Omega}_{s0}, \tag{26}%
\end{equation}

\begin{equation}
\left\langle 0\left\vert \Phi_{s}\right\vert 0\right\rangle =\overline{\Phi
}_{s0}=Diagonal\left(  1,1,1,-\frac{3}{2},-\frac{3}{2}\right)  \upsilon
_{\Phi0}, \tag{27}%
\end{equation}

\begin{equation}
\left\langle 0\left\vert \chi_{s}\right\vert 0\right\rangle ^{+}%
=\overline{\chi}_{s0}=\frac{\upsilon_{\chi0}}{\sqrt{2}}\left(
0,0,0,0,1\right)  , \tag{28}%
\end{equation}

\begin{equation}
\upsilon_{\Omega0}^{2}=\frac{\mu^{2}}{f},\;\ \ \ \ \ \ f\equiv\lambda
-\frac{15\eta^{2}}{15a+7b}-\frac{\varsigma^{2}}{\xi} \tag{29}%
\end{equation}

\begin{equation}
\upsilon_{\Phi0}^{2}=\frac{2\eta}{15a+7b}\upsilon_{\Omega0}^{2}, \tag{30}%
\end{equation}

\begin{equation}
\upsilon_{\chi0}^{2}=\frac{2\zeta}{\xi}\upsilon_{\Omega0}^{2}, \tag{31}%
\end{equation}
where $\lambda>15\eta^{2}/\left(  15a+7b\right)  +\zeta^{2}/\xi.$ Such $space$
in which the vacuum expectation values of the Higgs fields take the forms
$(25)-(31)$ is called $S-space.$ In $S-space$ the $S-SU(5)$ symmetry is
finally broken into $SU(3)\times U(1),$ $s-particles$ can get their masses of
the $SU(5)$ $GUT$ and form the $s-world$ described $SU(5)$ $GUT$, the
$V-SU(5)$ symmetry still holds, all $v-gauge\,\,bosons$ and $v-fermions$ are
massless. From $\left(  12\right)  $ and $\left(  13\right)  $ it can be
proved that all $v-Higgs$ $bosons$ can get their masses big enough. The masses
of the Higgs particles except the $\Phi_{s}-particles$ and the $\chi
_{s}-particles\ $are respectively%

\begin{equation}
m^{2}\left(  \Omega_{s}\right)  =2\mu^{2}, \tag{32}%
\end{equation}

\begin{equation}
m^{2}\left(  \Omega_{v}\right)  =\Lambda\upsilon_{\Omega0}^{2}-\mu^{2},
\tag{33}%
\end{equation}

\begin{equation}
m^{2}\left(  \Phi_{v}\right)  =\frac{1}{2}\alpha\upsilon_{\Omega0}^{2}
\tag{34}%
\end{equation}

\begin{equation}
m^{2}\left(  \chi_{v}\right)  =\beta\upsilon_{\Omega0}^{2}. \tag{35}%
\end{equation}
We can choose such parameters so that
\begin{equation}
m\left(  \Omega_{s}\right)  \simeq m\left(  \Omega_{v}\right)  \gg m\left(
\varphi_{s}\right)  \gg m\left(  \chi_{s}\right)  , \tag{36}%
\end{equation}
e.g., $m\left(  \Omega_{s}\right)  \sim10^{19}Gev,$ $m\left(  \varphi
_{s}\right)  \sim10^{15}Gev$ and $m\left(  \chi_{s}\right)  \sim10^{2}Gev$.
When radiative corrections are considered, the results above still
qualitatively hold. It is easily seen from $(32)-(35)$ that all real
components of $\Phi_{v}$ have the same mass $m\left(  \Phi_{v}\right)  $, all
real components of $\chi_{v}$ have the same mass $m\left(  \chi_{v}\right)  $
in $S-space.$

Because $s-matter$ and $v-matter$ are symmetric, $S-space$ and $V-space$ are
symmetric as well, when the subscripts $s$ and $v$ in formulas $(25)-(35)$ are
interchanged and the subscripts $S$ and $V$ in $(25)-(35)$ are interchanged as
well, i.e., $s\rightleftarrows v$ and $S\rightleftarrows V,$ the formulas
still hold water. Such space in which the vacuum expectation values of the
Higgs fields take the forms $(25)-(31)$ in which the subscripts $s$ and $v$
are interchanged is called $V-space.$

It is seen from the above mentioned that properties of space are determined by
breaking mode of the Higgs fields. Because breaking mode of the Higgs fields,
i.e., expectation values of the Higgs fields, can vary with temperature, the
properties of space will vary with temperature as well. We will see that when
temperature $T\longrightarrow T_{\max},$ $S-space$ can transform into
$V-space$, consequently $s-world$ can transform into $v-world$, and vice versa.

When the Higgs potentials do not considered, we have that the energy density
of the vacuum state $E_{0}=0^{[6]}$. Considering the potential energy
densities $V_{S}\left(  \varpi_{s},\varpi_{v}\right)  $ in $S-space$ to be,%
\begin{equation}
V_{S}\left(  \varpi_{s},\varpi_{v}\right)  =V_{s}\left(  \varpi_{s}\right)
+V_{0}+V_{v}\left(  \varpi_{v}\right)  +V_{sv}\left(  \varpi_{s},\varpi
_{v}\right)  . \tag{37}%
\end{equation}
Taking $V_{0}=-V_{s\min}$ at the 0-temperature, considering $V_{v\min
}=V_{sv\min}=0,$ we get energy densities of the vacuum state $E_{S0}$ in
$S-space$ at the 0-temperature to be
\begin{equation}
E_{S0}=E_{0}+V_{S\min}=E_{0}+\left(  V_{s}+V_{v}+V_{sv}\right)  _{\min}%
+V_{0}\text{\ }=0, \tag{38}%
\end{equation}
Similarly, we have%
\begin{equation}
E_{V0}=E_{0}+V_{V\min}=E_{0}+\left(  V_{s}+V_{v}+V_{sv}\right)  _{\min}%
+V_{0}\text{\ }=0\text{.} \tag{39}%
\end{equation}
\ \ 

\section{Evolution of space}

\subsection{Evolving equations}

Provided the cosmological principle holds, the metric tensor is the
Robertson-Walker metric which can be written%

\begin{equation}
(ds)^{2}=-\left(  dt\right)  ^{2}+R^{2}\left(  t\right)  \left\{
\frac{\left(  dr\right)  ^{2}}{1-kr^{2}}+\left(  rd\theta\right)  ^{2}+\left(
r\sin\theta d\varphi\right)  ^{2}\right\}  , \tag{40}%
\end{equation}
where $k$ is a real constant describing curve of space.

Matter in the universe may approximately be regarded as ideal gas to evenly
distribute in whole space. The energy-momentum tensor densities of the ideal
gas are
\begin{equation}
T_{As\mu\nu}=\left(  \rho_{As}+p_{As}\right)  U_{As\mu}U_{As\nu}+p_{As}%
g_{A\mu\nu}, \tag{41}%
\end{equation}%
\begin{equation}
T_{Av\mu\nu}=\left(  \rho_{Av}+p_{Av}\right)  U_{Av\mu}U_{Av\nu}+p_{Av}%
g_{A\mu\nu}, \tag{42}%
\end{equation}
where $A=S,V$, $\rho_{A}$ and $p_{A}$ are the density and pressure of a
medium, respectively, and $U_{A\mu}$ is a 4-velocity. Considering the
potential energy densities in $(15)-(16)$ and $(18)$, we can rewrite
$(41)-(42)$ as
\begin{equation}
T_{Ss\mu\nu\text{ }}=\left[  \widetilde{\rho}_{Ss}+\widetilde{p}_{Ss}\right]
U_{S\mu}U_{S\nu}+\widetilde{p}_{Ss}g_{S\mu\nu}, \tag{43}%
\end{equation}%
\begin{equation}
\widetilde{\rho}_{Ss}=\rho_{Ss}+\left(  V_{s}\left(  \varpi_{s}\right)
+V_{0}\right)  ,\;\ \ \ \widetilde{p}_{Ss}=p_{Ss}-\left(  V_{s}(\varpi
_{s})+V_{0}\right)  , \tag{44}%
\end{equation}%
\begin{equation}
T_{Sv\mu\nu\text{ }}=\left[  \widetilde{\rho}_{Sv}+\widetilde{p}_{Sv}\right]
U_{S\mu}U_{S\nu}+\widetilde{p}_{Sv}g_{S\mu\nu}, \tag{45}%
\end{equation}%
\begin{equation}
\widetilde{\rho}_{Sv}=\rho_{Sv}+V_{v}\left(  \varpi_{v}\right)
,\;\ \ \ \widetilde{p}_{Sv}=p_{Sv}-V_{v}(\varpi_{v}), \tag{46}%
\end{equation}%
\begin{equation}
T_{Vs\mu\nu\text{ }}=\left[  \widetilde{\rho}_{Vs}+\widetilde{p}_{Vs}\right]
U_{V\mu}U_{V\nu}+\widetilde{p}_{Vs}g_{V\mu\nu}, \tag{47}%
\end{equation}%
\begin{equation}
\widetilde{\rho}_{Vs}=\rho_{Vs}+V_{s}\left(  \varpi_{s}\right)
,\;\ \ \ \widetilde{p}_{Vs}=p_{Vs}-V_{s}(\varpi_{s}), \tag{48}%
\end{equation}%
\begin{equation}
T_{Vv\mu\nu\text{ }}=\left[  \widetilde{\rho}_{Vv}+\widetilde{p}_{Vv}\right]
U_{V\mu}U_{V\nu}+\widetilde{p}_{Vv}g_{V\mu\nu}, \tag{49}%
\end{equation}%
\begin{equation}
\widetilde{\rho}_{Vv}=\rho_{Vv}+\left(  V_{v}\left(  \varpi_{v}\right)
+V_{0}\right)  ,\;\ \ \ \widetilde{p}_{Vv}=p_{Vv}-\left(  V_{v}(\varpi
_{v})+V_{0}\right)  , \tag{50}%
\end{equation}
where $\rho_{Aa}$ and $p_{Aa}$ $(a=s,v)$ are respectively the motional mass
and the momentum current densities corresponding to $\mathcal{L}_{sm}$ and
$\mathcal{L}_{vm}$ in $(15),$ $(16)$ and $(18)$. The difference between
$\rho_{Ss}$ and $\rho_{Vs}$ is that $\rho_{Ss}$ is given common matter density
in $S-space$ and $\rho_{Vs}$ is the dark energy density in $V-space$.

Substituting $(43)-(46)$ and the Robertson-Walker metric into $(19)$ and
considering $U_{S\mu}=(1,0,0,0)$ which implies that a medium move only as
expanding of the universe, we get the generalized Freedman equations%

\begin{align}
\frac{\overset{\cdot}{R}_{S}^{2}(t_{S})}{R_{S}^{2}\left(  t_{S}\right)
}+\frac{k_{S}}{R_{S}^{2}\left(  t_{S}\right)  }  &  =\eta\left[
\widetilde{\rho}_{s}-\widetilde{\rho}_{v}\right] \nonumber\\
&  =\eta\left[  \left(  \rho_{s}+V_{s}(\varpi_{s})+V_{0}\right)  -\left(
\rho_{v}+V_{v}(\varpi_{v})\right)  \right]  , \tag{51}%
\end{align}
where $\eta\equiv8\pi G/3,$%
\begin{align}
\frac{\overset{\cdot\cdot}{R}_{S}(t_{S})}{R_{S}\left(  t_{S}\right)  }  &
=-\frac{1}{2}\eta\left[  \left(  \widetilde{\rho}_{s}+3\widetilde{p}%
_{s}\right)  -\left(  \widetilde{\rho}_{v}+3\widetilde{p}_{v}\right)  \right]
\nonumber\\
&  =-\frac{1}{2}\eta\left[  \left(  \rho_{s}+3p_{s}\right)  -2\left(
V_{s}(\varpi_{s})+V_{0}\right)  -\left(  \rho_{v}+3p_{v}\right)
+2V_{v}(\varpi_{v})\right]  . \tag{52}%
\end{align}
Analogously, from $(46)-(49)$, $(40)$ and $(20)$, we get
\begin{align}
\frac{\overset{\cdot}{R}_{V}^{2}(t_{V})}{R_{V}^{2}\left(  t_{V}\right)
}+\frac{k_{V}}{R_{V}^{2}\left(  t_{V}\right)  }  &  =\eta\left[
\widetilde{\rho}_{v}-\widetilde{\rho}_{s}\right] \nonumber\\
&  =\eta\left[  \rho_{v}+V_{v}(\varpi_{v})+V_{0}-\rho_{s}-V_{s}(\varpi
_{s})\right]  , \tag{53}%
\end{align}%
\begin{align}
\frac{\overset{\cdot\cdot}{R}_{V}(t_{V})}{R_{V}\left(  t_{V}\right)  }  &
=-\frac{1}{2}\eta\left[  \left(  \widetilde{\rho}_{v}+3\widetilde{p}%
_{v}\right)  -\left(  \widetilde{\rho}_{s}+3\widetilde{p}_{s}\right)  \right]
\nonumber\\
&  =-\frac{1}{2}\eta\left[  \left(  \rho_{v}+3p_{v}\right)  -2\left(
V_{v}(\varpi_{v})+V_{0}\right)  -\left(  \rho_{s}+3p_{s}\right)
+2V_{s}(\varpi_{s})\right]  . \tag{54}%
\end{align}

Because of the coupling term $V_{sv}$, the expectation values of the Higgs
fields can only take the following forms. $\langle\omega_{s}\rangle\neq0$ and
$\langle\omega_{v}\rangle=0$ or $\langle\omega_{s}\rangle=0$ and
$\langle\omega_{v}\rangle\neq0$ when temperature is low. Hence space is only
$S-space$ in which only $(51)-(52)$ applies, or only $V-space$ in which only
$(53)-(54)$ applies. It is possible that $\langle\omega_{s}\rangle=$
$\langle\omega_{v}\rangle=0$ when temperature is high enough ($T\sim T_{\max}%
$). Such space in which $\langle\omega_{s}\rangle=$ $\langle\omega_{v}%
\rangle=0$ is called transiting space. In the case, $s-matter$ and $v-matter$
is completely symmetric, and both $(51)-(52)$ and $(53)-(54)$ can describe
evolution of transiting space, i.e., transiting space can transform into
$S-space$ or $V-space.$ This is because a physical quantity or an equation has
its meanings only when it is relative to a definite physics vacuum.

For example, let a state be the vacuum of $S-space,$ i.e., $\rho_{a}%
=p_{a}=V_{v}(\varpi_{v})=0,$ and $V_{s}(\varpi_{s})=-\mu^{2}/4f.$ Substituting
the parameters into $(51)-(52)$ and taking $k_{S}=0$, we get
\[
\overset{\cdot\cdot}{R}_{S}\left(  t_{S}\right)  =\overset{\cdot}{R}%
_{S}\left(  t_{S}\right)  =0,\;\ \ R_{S}\left(  t_{S}\right)  =R_{S}\left(
0\right)  ,
\]
i.e., $S-space$ is invariant. Oppositely, if substituting the parameters into
$(53)-(54)$ and taking $k_{V}=0,$ we get%
\[
\frac{\overset{\cdot\cdot}{R}_{V}(t_{V})}{R_{V}\left(  t_{V}\right)  }=\eta
V_{0},\text{ \ }\frac{\overset{\cdot}{R}_{V}^{2}(t_{V})}{R_{V}^{2}\left(
t_{V}\right)  }=\eta V_{0},\text{ \ }R_{V}\left(  t_{V}\right)  =R_{V}\left(
0\right)  \exp t_{V}\sqrt{\eta V_{0}}\text{\ \ or }0\text{.}%
\]
This result is obviously difficult to understand it. Appearance of the result
is because the state above is not the ground state of $V-space.$ There are
similar examples in physics. For example, for Higgs field with single
constituent, if the state were regarded as the ground state in which
$\langle\varphi\rangle=0$ $(V\left(  0\right)  =V_{\max}\left(  \varphi
\right)  ),$ the `$mass$' of $\varphi-particle$ would be a virtual and thereby
$\varphi-particle$ would move in a greater velocity than photon velocity. Of
course, this is not right.

Because of the coupling term $(13)$, the direct transformation, $\langle
\omega_{s}\rangle\neq0$ and $\langle\omega_{v}\rangle=0\longrightarrow
\langle\omega_{s}\rangle=0$ and $\langle\omega_{v}\rangle\neq0$, is
impossible, i.e., $S-space$ cannot directly transform into $V-space.$ But the
transformation
\begin{equation}
\langle\omega_{s}\rangle\neq0,\text{ }\langle\omega_{v}\rangle
=0\longrightarrow\langle\omega_{s}\rangle=\langle\omega_{v}\rangle
=0\longrightarrow\langle\omega_{s}\rangle=0\text{, }\langle\omega_{v}%
\rangle\neq0 \tag{55}%
\end{equation}
is possible, i.e., $S-space$ can transform into $V-space$ via transiting space
as temperature rises to the highest temperature (see the following).

Considering $(38)$, $\widetilde{\rho}_{a}=\rho_{a},$ $\widetilde{p}_{a}%
=p_{a},$ $a=s,v,$ from $(51)-(52)$ we obtain%
\begin{equation}
d\left[  \left(  \rho_{s}-\rho_{v}\right)  R_{S}^{3}\right]  =-\left(
p_{s}-p_{v}\right)  dR_{S}^{3}. \tag{56}%
\end{equation}
When temperature is low so that the pressures $p_{s}$ and $p_{v}$ may be
ignored. Thus $\left(  56\right)  $ become%
\begin{equation}
\ \left(  \rho_{s}-\rho_{v}\right)  R_{S}^{3}=C_{S},\text{\ \ \ } \tag{57}%
\end{equation}
in $S-space$, here $C_{S}$ is a constant. In contrast to the conventional
theory, it is possible $C_{S}>0,$ $C_{S}=0$ or $C_{S}<0.$ Analogously, we
have
\begin{equation}
\ \left(  \rho_{v}-\rho_{s}\right)  R_{V}^{3}=C_{V},\text{\ \ \ } \tag{58}%
\end{equation}
in $V-space$.

\bigskip For photon-like gases in heat balance, $p_{a}=\rho_{a}/3,$ we have
\begin{equation}
\left(  \rho_{s}-\rho_{v}\right)  R_{S}^{4}=C_{S}^{\prime}, \tag{59}%
\end{equation}
in $S-space,$ and
\begin{equation}
\ \left(  \rho_{v}-\rho_{s}\right)  R_{V}^{4}=C_{V}^{\prime},\text{\ \ \ }
\tag{60}%
\end{equation}
in $V-space$.

When the approximations $\left(  57\right)  $ and $(58)$ are taken$,$ i.e.,
temperature is low, the equations $\left(  51\right)  -\left(  54\right)  $
become%
\begin{equation}
\overset{\cdot}{R}_{S}^{2}(t_{S})=-k_{S}+\eta\left(  \rho_{s}-\rho_{v}\right)
R_{S}^{2}, \tag{61}%
\end{equation}%
\begin{equation}
\overset{\cdot\cdot}{R}_{S}(t_{S})=-\frac{\eta}{2}\left(  \rho_{s}-\rho
_{v}\right)  R_{S}, \tag{62}%
\end{equation}%
\begin{equation}
\overset{\cdot}{R}_{V}^{2}(t_{V})=-k_{V}+\eta\left(  \rho_{v}-\rho_{s}\right)
R_{V}^{2}, \tag{63}%
\end{equation}%
\begin{equation}
\overset{\cdot\cdot}{R}_{V}(t_{V})=-\frac{\eta}{2}\left(  \rho_{v}-\rho
_{s}\right)  R_{V}. \tag{64}%
\end{equation}

\subsection{Two sorts of temperature}

Because there is no interaction except the repulsion between $s-matter$ and
$v-matter$ and the masses of the Higgs-particles are all big enough at low
temperature, the Higgs bosons are hardly produced. Consequently there is no
thermal equilibrium between the $v-particles$ and the $s-particles.$ Thus at
low temperature, we should use two sorts of temperature $T_{v}$ and $T_{s}$
respectively to describe the thermal equilibrium of $v-matter$ and the thermal
equilibrium of $s-matter.$ In general case, $T_{v}\neq T_{s}$. When
temperature $T_{s}$ is so high that $\langle\Omega_{s}\rangle\longrightarrow0$
in $S-space$, the masses of Higgs bosons tend to zero. On the other hand,
because $\langle\Omega_{v}\rangle=0$ in $s-space,$ $m\left(  \Omega
_{v}\right)  ,$ $m\left(  \Phi_{v}\right)  $ and $m\left(  \chi_{v}\right)  $
tend to zero as $\langle\Omega_{s}\rangle$ tends to zero as well. Consequently
$\Omega_{s},$ $\Phi_{s}$ and $\chi_{s}$ can be enormously produced and easily
transform into $\Omega_{v},$ $\Phi_{v}$ and $\chi_{v}$ by the couplings in
$(13).$ Other $v-particles$ can be easily produced by the couplings of
$V-SU(5)$ as well$.$ Consequently there is thermal equilibrium between the
$v-particles$ and the $s-particles$ when $T_{s}\sim T_{v}\sim T_{\max}$. In
the case, contraction of space will stop and inflating must occur. Thus the
temperature $T>T_{\max}$ cannot appear in fact.

\subsection{Temperature effect}

The influences of finite temperature on the Higgs potential in the present
model are consistent with the conventional theory. For
\[
V(\Omega_{s})=-\frac{\mu^{2}}{2}\Omega_{s}^{2}+\frac{\lambda}{4}\Omega_{s}%
^{4},
\]
to ignore the terms proportional $\lambda^{n},$ $n>1$, the effective potential
approximate to 1-loop at finite temperature and in flat space is$^{\left[
9\right]  }$
\begin{equation}
V_{eff}^{(1)}\left(  \overline{\Omega}_{s},T_{s}\right)  =-\left(  \frac
{\mu^{2}}{2}-\frac{\lambda T_{s}^{2}}{8}\right)  \overline{\Omega}_{s}%
^{2}+\frac{\lambda}{4}\overline{\Omega}_{s}^{4}-\frac{\pi^{2}}{90}T_{s}%
^{4}+\frac{\mu^{2}}{24}T_{s}^{2}, \tag{65}%
\end{equation}
It is seen from the term $\lambda T_{s}^{2}\overline{\Omega}_{s}^{2}/8$ that
the expectation value $\overline{\Omega}_{s}$ can be strikingly altered at
high temperature. When $T_{s}>\sqrt{4\mu^{2}/\lambda},$ $\overline{\Omega}%
_{s}=0.$ For
\[
V(\Phi_{s})=-\frac{1}{2}\kappa\Omega_{s}^{2}Tr\Phi_{s}^{2}+\frac{1}{4}%
a(Tr\Phi_{s}^{2})^{2}+\frac{1}{2}bTr\Phi_{s}^{4},
\]
ignoring the contributions of Higgs fields and fermion fields with one loop
correction and only considering the contribution of the gauge fields with loop
correction, when $\overline{\varphi}_{s}\ll kT$, $k$ is the Boltzmann
constant, the effective potential approximate to 1-loop at finite temperature
and in flat space is$^{\left[  10\right]  }$
\begin{equation}
V_{eff}^{(1)}\left(  \overline{\Phi}_{s},T_{s}\right)  =V(\overline{\Phi}%
_{s})+B\overline{\varphi}_{s}^{4}\left(  \ln\frac{\overline{\varphi}_{s}^{2}%
}{\sigma^{2}}-\frac{25}{6}\right)  +\frac{75}{16}\left(  kgT\right)
^{2}\overline{\varphi}_{s}^{2}-\frac{\pi^{2}}{15}\left(  kT\right)  ^{4},
\tag{66}%
\end{equation}
where $B=\left(  5625/1024\pi^{2}\right)  g^{4},$ $\sigma=\left\langle
\varphi_{s}\right\rangle \sim10^{15}Gev.$

Taking the place of the subscript $s$ by $v$ in $(65)$ and $(66)$, we get
$V_{eff}^{(1)}\left(  \overline{\Omega}_{v},T_{v}\right)  $ and $V_{eff}%
^{(1)}\left(  \overline{\Phi}_{v},T_{v}\right)  .$ The effective potential
approximate to 1-loop at finite temperature and in flat space can be written
as%
\begin{equation}
V_{eff}^{(1)}\left(  \varpi_{s},T_{s},\varpi_{v},T_{v}\right)  =(V_{eff}%
^{(1)}\left(  \varpi_{s},T_{s}\right)  +V_{0})+V_{eff}^{(1)}\left(  \varpi
_{v},T_{v}\right)  +V_{sv}\left(  \varpi_{s},\varpi_{v}\right)  . \tag{67}%
\end{equation}

For short, we only consider influence of temperature on $\Omega_{s}$ and
$\Omega_{v}$ and ignore the couplings of $\Omega_{s}$ and $\Omega_{v}$ with
other fields. Considering $V_{Seff}^{(1)}$ in $S-space$ and $\Lambda>\lambda,$
ignoring the terms irrelative to $\Omega_{s}$ and $\Omega_{v},$ we have
\begin{align}
V_{Seff}^{(1)}\left(  \overline{\Omega}_{s},T_{s},\overline{\Omega}_{v}%
,T_{v}\right)   &  =\left[  -\frac{1}{2}\left(  \mu^{2}-\lambda T_{s}%
^{2}/4\right)  \overline{\Omega}_{s}^{2}+\frac{1}{4}\lambda\overline{\Omega
}_{s}^{4}+V_{0}\right] \nonumber\\
&  +\left[  -\frac{1}{2}\left(  \mu^{2}-\lambda T_{v}^{2}/4\right)
\overline{\Omega}_{v}^{2}+\frac{1}{4}\lambda\overline{\Omega}_{v}^{4}\right]
+\frac{1}{2}\Lambda\overline{\Omega}_{s}^{2}\overline{\Omega}_{v}^{2}.
\tag{68}%
\end{align}
From $(66)$ and $(68)$ we see the absolute values of the expectation values of
the Higgs fields will decrease as the temperature $T_{s}$ rises, consequently
$V_{eff}^{(1)}\left(  \varpi_{s},T_{s},\varpi_{v},T_{v}\right)  $ must
increase. From $(68)$ we get when $\mu^{2}-\lambda T_{s}^{2}/4\geqslant0$
\begin{equation}
m^{2}\left(  \Omega_{s},T_{s}\right)  =2\left(  \mu^{2}-\frac{1}{4}\lambda
T_{s}^{2}\right)  \geqslant0, \tag{69}%
\end{equation}
and when
\[
\frac{\Lambda}{\lambda}\left(  \mu^{2}-\frac{1}{4}\lambda T_{s}^{2}\right)
-\left(  \mu^{2}-\frac{\lambda}{4}T_{v}^{2}\right)  \geqslant0,
\]%
\begin{equation}
m^{2}\left(  \Omega_{v},T_{s},T_{v}\right)  =\frac{\Lambda}{\lambda}\left(
\mu^{2}-\frac{1}{4}\lambda T_{s}^{2}\right)  -\left(  \mu^{2}-\frac{\lambda
}{4}T_{v}^{2}\right)  \geqslant0, \tag{70}%
\end{equation}%
\begin{equation}
\overline{\Omega}_{s}^{2}\left(  T_{s}\right)  =\left(  \mu^{2}-\lambda
T_{s}^{2}/4\right)  /\lambda\text{, \ \ \ \ \ }\overline{\Omega}_{v}\left(
T_{v}\right)  =0. \tag{71}%
\end{equation}
Because $\Lambda>\lambda,$ when $T_{s}\sim T_{v}\sim0,$ $m^{2}\left(
\Omega_{v},T_{s},T_{v}\right)  >0.$ As $T_{s}$ and $T_{v}$ rise$,$
$m^{2}\left(  \Omega_{s},T_{s}\right)  $ and $m^{2}\left(  \Omega_{v}%
,T_{s},T_{v}\right)  $ will decrease. From $(68)$-$\left(  70\right)  $ we see
it is necessary that when $\mu^{2}-\lambda T_{s}^{2}/4\longrightarrow0,$
$m\left(  \Omega_{v}\right)  \sim m\left(  \Omega_{s}\right)  \longrightarrow
0$. In the case, $s-Higgs$ particles can easily transform into $v-Higgs$
particles by the coupling $(13)$ so that the thermal equilibrium between
$s-matter$ and $v-matter$ will come into being, $T_{v}\sim T_{s}$ and
$\rho_{s}\simeq\rho_{v}.$ Thus contraction of $s-space$ must stop and
inflation of $v-space$ will occur (see following).

In fact, when the energy density $\rho_{s\text{ }}-\rho_{v\text{ }}$ is large
and $T_{v}<T_{s}\precsim T_{\max}$, it is possible that $S-space$ can contract
so swiftly that the temperature rises to $T_{\max}$ before the potential
reduces to
\begin{align*}
V_{Veff\min}^{(1)}\left(  \varpi_{s},\varpi_{v},T_{\max}\right)   &
=V_{seff\min}^{(1)}\left(  \varpi_{s},T_{\max}\right)  +(V_{0}+V_{veff\min
}^{(1)}\left(  \varpi_{v},T_{v}\right)  )+V_{sv\min}\left(  \varpi_{s}%
,\varpi_{v}\right) \\
&  =V_{0}+V_{veff\min}^{(1)}\left(  \varpi_{v},T_{v}\right)  >0,
\end{align*}
here $V_{veff\min}^{(1)}$ is the potential energy corresponding to $\left(
\Lambda/\lambda\right)  \left(  \mu^{2}-\frac{1}{4}\lambda T_{s}^{2}\right)
-\left(  \mu^{2}-T_{v}^{2}/4\right)  <0.$ In this case, we have still
$\langle\omega_{s}\rangle=\langle\omega_{v}\rangle=0,$ $m\left(  \omega
_{s}\right)  =m\left(  \omega_{s}\right)  =0,$ $s-Higgs$ particles and
$v-Higgs$ particles can transform from one to other, consequently the thermal
equilibrium can come into being by the couplings in $(13),$ i.e., $T_{s}%
=T_{v}$ and $\rho_{s}=\rho_{v}$. Thus the inflation of the universe still
occurs. The state with $V_{S\min}\left(  \varpi_{s},\varpi_{v},T_{s}%
,T_{v}\right)  $ is a steady state at $T_{s}$. The process is called
\textbf{super-heating process }in which the temperature of a system rises to
$T$ before the potential energy $V(T)$ decrease to corresponding $V_{\min}%
(T)$. The contracting process is a \textbf{super-heating process}.

Considering $\Phi_{s},$ $\chi_{s},$ $\Phi_{v}$ and $\chi_{v},$ the conclusions
above still hold water.

\subsection{Contraction of space, the highest temperature and transiting
space}

\subsubsection{Contraction of space and the highest temperature}

In contrast with the conventional cosmological model, the present model can
explain the \textquotedblleft big bang\textquotedblright\ of the universe, but
there is no\ singularity.

In $S-space$, $V-SU(5)$ does not break so that all $v-gauge$ $bosons$ and
$v-fermions$ are massless and they form $V-SU(5)$ colour single states with
their masses in low temperature, and $S-SU(5)$ breaks into $S-SU(3)\times
U(1)$ so that $s-particles$ gets their masses and can form $s-galaxies$. In
$S-space,$ let $\widetilde{\rho}_{s}-\widetilde{\rho}_{v}>0,$ $\left(
\widetilde{\rho}_{s}+3\widetilde{p}_{s}\right)  -\left(  \widetilde{\rho}%
_{v}+3\widetilde{p}_{v}\right)  >0$ and $k=1.$ From $\left(  51\right)
$-$\left(  52\right)  $ or $\left(  61\right)  $-$\left(  62\right)  $ we see
that $S-space$ will contract, consequently temperature $T_{s}$ will rise$,$
and $V_{s}$ can increase as $T_{s}$ rises since the expectation values of the
$S-Higgs\,\ fields$ can vary with $T_{s}$.

From $(69)$-$(70)$ we see when $\lambda T_{s}^{2}/4\sim\mu^{2}$, $m(\Omega
_{s})\sim m(\Omega_{v})\sim0.$ In the case, the masses of $\Omega_{s}$,
$\varphi_{s}$ and $\chi_{s\text{ }}$tend to zero, hence the $s-fermions$ and
the $s-gauge$ bosons can easily transform into $\chi_{s}$, $\varphi_{s}$ and
$\Omega_{s},$ and $\Omega_{s}$, $\varphi_{s}$ and $\chi_{s}$ can easily
transform into $\Omega_{v}$, $\varphi_{v}$ and $\chi_{v}$ by the couplings in
$(13)$. As a consequence, we have
\begin{equation}
T_{s}\sim T_{v},\text{ \ \ }\rho_{s}\sim\rho_{v}.\text{ } \tag{72}%
\end{equation}
In the case, $p_{s}=\rho_{s}/3$ and $p_{v}=\rho_{v}/3$, considering $k=1$ for
contracting $S-space$ , from $(51)-(52)$ we get
\begin{align}
\overset{\cdot}{R}_{S}^{2}\left(  t_{S}\right)   &  =-1+\eta V_{0}R_{S}%
^{2}\left(  t_{S}\right)  ,\text{ \ \ }\overset{\cdot}{R}_{S}\left(
t_{S}\right)  =-\sqrt{-1+\eta V_{0}R_{S}^{2}\left(  t_{S}\right)
},\nonumber\\
\overset{\cdot\cdot}{R}_{S}\left(  t_{S}\right)   &  =\eta V_{0}R_{S}\left(
t_{S}\right)  >0. \tag{73}%
\end{align}
It is seen that in the case, $S-space$ will contract with a deceleration and
when
\begin{equation}
R_{S}\left(  t_{S}\right)  \equiv R_{S}\left(  t_{SF}\right)  \equiv R_{\min
}=\sqrt{\frac{1}{\eta V_{0}}},\text{ \ \ \ \ }\overset{\cdot}{R}_{S}\left(
t_{SF}\right)  =0. \tag{74}%
\end{equation}
Consequently $R_{S}\left(  t_{SF}\right)  \equiv R_{\min}$ cannot continue to
reduce, $t_{SF}$ is the last moment of the $S-world$ and the initial moment of
the $V-world$, and $T_{s}(t_{SF})$ and $\rho_{s}=\rho_{v}=\rho_{\max}$ cannot
continue to increase. The temperature $T_{s}(t_{SL})\equiv T_{\max}$
corresponding to $R_{\min}$ is the \textbf{highest temperature}, $\rho_{\max}$
is the largest density and $R_{\min}$ is the minimum scale. Consequently, we
see that according to the present model, there are the $R_{\min}$, $T_{\max}$
and $\rho_{\max},$ and considering the case described by $\left(  72\right)
-(73)$, there is no singularity in any case.

Considering $\lambda\sim g^{4}$, $g\sim4\pi/45$ for $SU(5)$ and $m(\Omega
_{s})=\sqrt{2}\mu$, from $(69)$ we see that $T_{\max}$ can be estimated,
\begin{equation}
T_{\max}\sim\frac{2\mu}{\sqrt{\lambda}}\sim\frac{2\mu}{g^{2}}\sim\frac
{\sqrt{2}m(\Omega_{s})}{4\pi/45}=5m(\Omega_{s}). \tag{75}%
\end{equation}

As is well known, there is the singular point in the universe according to the
conventional cosmological theory. This is because there is only one sort of
matter and only one sort of space in the conventional cosmological theory. To
consider the effective vacuum energy density $\rho_{vac}$, the evolving
equations of the universe are%
\[
\frac{\overset{\cdot}{R}^{2}(t)}{R^{2}\left(  t\right)  }+\frac{k}%
{R^{2}\left(  t\right)  }=\frac{8\pi G}{3}(\rho+\rho_{vac}),
\]%
\[
\frac{\overset{\cdot\cdot}{R}(t)}{R\left(  t\right)  }=-\frac{8\pi G}{3}%
(\rho-\rho_{vac}).
\]
Because $\rho_{vac}$ is a constant, we see that provided $\rho$ is large
enough, there must be singular point in the universe. Of course, in fact, the
cosmological theory assumes now $\rho\left(  t_{0}\right)  -\rho_{vac}<0$ and
thereby explains expansion of the universe with an acceleration. But expansion
of the universe must begin from a singular point.

In contrast with the conventional expansive model, there are the two sorts of
matter and the two sorts of space$.$ One condition of the space-time
singularity theorems is violated, consequently there is no the space-time
singularity. As is mentioned above, no matter how large $\rho_{s}$ or
$\rho_{v}$ is, when $T_{s}$ or $T_{v}\sim T_{\max}$, the universe can no
longer contract, and inflation must occur. Hence there is no singular point
and there is the highest temperature $T_{\max}$ in the universe$.$

\subsubsection{Transiting space and selection of $k$}

When $T_{s}$ $\sim T_{v}\sim T_{\max},$ $\varpi_{s}=\varpi_{v}=0$ and
$\rho_{s}\sim\rho_{v}\sim\rho_{\max}.$ In the case, $s-matter$ and $v-matter$
is completely symmetric and space cannot be regarded as $S-space$ neither
$V-space.$ space in the state may be called transiting space ($T-space$).
$T-space$ is not stable and will continue to evolve as $V-space$ or $S-space.$
$R_{S}\equiv R_{\min}$ implies the $S-world$ have ended and a new world will
begin. $k=1$ for the contracting $S-space.$ It is not necessary $k=1$ for the
new world and new space. We take $k=-1.$In fact$,$ $k=1,0,$ or $-1$ is not
important for $T-space.$

\subsection{Inflating of space}

From $(73)-(74)$ we see that when $R_{S}\equiv R_{\min}$, space begin to
expand. Let transiting space transforms into $V-space$. From $(53)-(54)$ we
get the evolving equations of $V-space$ in the case,%
\begin{align}
\overset{\cdot}{R}_{V}^{2}\left(  t_{V}\right)   &  =1+\eta V_{0}R_{V}%
^{2}\left(  tV\right)  ,\text{ \ \ }\overset{\cdot}{R}_{V}\left(
t_{V}\right)  =\sqrt{1+\eta V_{0}R_{V}^{2}\left(  t_{V}\right)  }\tag{76}\\
\overset{\cdot\cdot}{R}_{V}\left(  t_{V}\right)   &  =\eta V_{0}R_{V}\left(
t_{V}\right)  . \tag{77}%
\end{align}
$(76)-(77)$ show that $V-space$ expands with an acceleration. Ignoring 1, from
$(76)-(77)$ we have
\begin{align}
\overset{\cdot}{R}_{V}\left(  t_{V}\right)  /R_{V}\left(  t_{V}\right)   &
=\sqrt{\eta V_{0}},\text{ \ \ }R_{V}\left(  t_{V}\right) \nonumber\\
&  =R_{V}\left(  t_{V1}\right)  \exp\sqrt{\eta V_{0}}\left(  t_{V}%
-t_{V1}\right)  =R_{V}\left(  0\right)  \exp Ht_{V}=R_{V}\left(  0\right)
\exp aT_{\max}^{2}t_{V},. \tag{78}%
\end{align}
where $H\equiv\sqrt{\eta V_{0}}=aT_{\max}^{2}$ is the Hubble\ constant at
$t_{V1},$ $a\equiv\sqrt{\eta}(\lambda/8\sqrt{f}).$

The temperature will strikingly decrease in the process of inflation. But the
potential energy $V\left(  \varpi_{s}\sim\varpi_{v}\sim0\right)  $ cannot
decrease to $V_{\min}\left(  \varpi_{s},\varpi_{v},T_{v}\right)  $ at once.
This is a \textbf{super-cooling process. } From $(78)$ we can get the
expectant results by suitably choosing the parameters in $(78)$.

From $(78)$ we can get the expectant results by suitably choosing the
parameters in $(78)$. For example, taking $\lambda/8\sqrt{f}\sim1,$
considering $\langle\omega_{s}\rangle=\langle\omega_{v}\rangle=0$ and
consequently the $SU(5)$ symmetry strictly holds water at $T_{\max}$, we can
take $T_{\max}$ to be the temperature corresponding to $GUT,$ i.e., $T_{\max
}\sim m\left(  \Omega_{s}\right)  \sim10^{15}Gev$. Thus we can get the same
result as the given inflation model. If the existing time of the
\textbf{super-cooling state }(i.e., the\textbf{ }sub-steady state with
$V\left(  \varpi_{s}\sim\varpi_{v}\sim0\right)  \sim V_{0})$ is $10^{-33}s$
and to take $T_{\max}\sim10^{15}Gev$, $H^{-1}\sim10^{-35}s$ $,$ $R_{S}\left(
0\right)  $ will increase $exp100\sim10^{43}$ times. The result is consistent
with the Guth's inflation model$^{[11]}$.

\subsection{The phase transition of the vacuum and the reheating process}

After inflation, the temperature must strikingly descend, consequently the
phase transition of the vacuum must occur. In\textbf{ }$T-space$, $S-SU(5)$
and $V-SU(5)$ symmetry strictly hold water and $s-matter$ and $v-matter$ are
strictly symmetric. Transiting space can transform into either $S-space$ or
$V-space$ by the phase transition of the vacuum. Let transiting
space\ transform into $V-space$ and $k=-1$ for $V-space$. Thus the phase
transition of the vacuum is follows,
\begin{align}
\varpi_{v}(T_{\max})  &  =0\longrightarrow\varpi_{v}(T_{v}\sim0)=\varpi
_{v0},\text{ \ \ \ }\varpi_{s}(T_{\max})=0\longrightarrow\varpi_{s}(T_{s}%
\sim0)=0,\nonumber\\
V_{v}\left(  \varpi_{v}=0\right)  +V_{0}  &  =V_{0}\longrightarrow
V_{v}\left(  \varpi_{v}=\varpi_{v0}\right)  +V_{0}=0,\text{\ \ \ \ }%
V_{s}\left(  \varpi_{s}=0\right)  \longrightarrow V_{s}\left(  \varpi
_{s}=0\right)  =0. \tag{79}%
\end{align}
After the transition, $V_{v}\left(  \varpi_{v}=0\right)  -V_{v}\left(
\varpi_{v}=\varpi_{v0}\right)  =V_{0}$ must directly transform into the energy
of $v-Higgs$ particles. $V-Higgs$ particles can fast decay into $v-gauge$
bosons and $v-fermions$ by $V-SU(5)$ couplings. On the other hand, because of
$m\left(  \Omega_{v}\right)  \sim m\left(  \Omega_{s}\right)  \gg m\left(
\varphi_{s}\right)  \gg m\left(  \chi_{s}\right)  $ and the couplings in
$(13),$ $v-Higgs$ particles can transform into $\Omega_{s},$ $\varphi_{s}$ and
$\chi_{s}$ as well$.$ $\Omega_{s},$ $\varphi_{s}$ and $\chi_{s}$ will fast
decay into $s-gauge$ bosons and $s-fermions.$ Let $\alpha V_{0}$ transforms
the $v-energy$, then $(1-\alpha)V_{0}$ transforms the $s-energy$ and it is
necessary $\alpha>(1-\alpha).$ Let before the transition, $\rho_{v}^{\prime
}=\rho_{s}^{\prime}.$ Thus after inflation, it is necessary that%
\begin{equation}
\rho_{v}=\rho_{v}^{\prime}+\alpha V_{0}>\rho_{s}=\rho_{s}^{\prime}%
+(1-\alpha)V_{0},. \tag{80}%
\end{equation}
This is the reheating process. After the reheating process, the temperatures
$T_{s}$ and $T_{v}$ rise higher, $v-particles$ get their masses and form
$v-atom$ and $v-galaxies,$ but $s-gauge$ bosons and elementary $s-fermions$
are still massless since $S-SU(5)$ is not broken. The $s-particles$ will form
$S-SU(5)$ colour single states analogous to the $SU(3)$ glue-balls whose
masses are not zero in low temperature.

After the phasic transition, $v-particles$ form the $v-world,$ $S-SU(5)$
colour single states are identified as dark energy in $v-world$. After the
reheating process, considering $k=-1$, in international unit $(53)-(54)$
become
\begin{align}
\overset{\cdot}{R}_{V}^{2}(t_{V})  &  =c^{2}+\eta\left[  \rho_{vm}-\rho
_{sm}+\rho_{v\gamma}\right]  R_{V}^{2}\left(  t_{V}\right)  ,\text{
\ or}\nonumber\\
\overset{\cdot}{a}^{2}  &  =\frac{c^{2}}{R_{0}^{2}}\left[  1-\frac{H_{0}%
^{2}R_{0}^{2}}{c^{2}}\left(  \frac{\Omega_{m0}}{a}-\frac{\Omega_{v\gamma0}%
}{a^{2}}\right)  \right]  , \tag{81}%
\end{align}%
\begin{equation}
\overset{\cdot\cdot}{R}_{V}(t_{V})=-\frac{1}{2}\eta\left(  \rho_{vm}-\rho
_{sm}+2\rho_{v\gamma}\right)  ,\text{ \ or\ \ }\overset{\cdot\cdot}{a}%
=\frac{1}{2}H_{0}^{2}\left(  \frac{\Omega_{m0}}{a^{2}}-2\frac{\Omega
_{v\gamma0}}{a^{3}}\right)  . \tag{82}%
\end{equation}
where $a=R\left(  t\right)  /R_{0},$ $\overset{\cdot}{a}_{0}^{2}=H_{0}%
^{2}=\eta\rho_{c}$, $\Omega_{m0}=-\left(  \rho_{vm0}-\rho_{sm0}\right)
/\rho_{c}$, $\Omega_{v\gamma0}=\rho_{v\gamma0}/\rho_{c},$ and we consider
$\rho_{s\gamma}=0$ in $V-space$ and ignore $p_{vm0}$ and $p_{sm0}.$ From
$(81)$ we have $H_{0}^{2}R_{0}^{2}/c^{2}=1/\left(  \Omega_{m0}-\Omega
_{v\gamma0}\right)  =1/\left(  1+\Omega_{0}\right)  ,$ $\Omega_{0}=\Omega
_{m0}-\Omega_{v\gamma0}.$ Thus $(81)$ become
\begin{equation}
\overset{\cdot}{a}^{2}=H_{0}^{2}\left(  1+\Omega_{0}\right)  \left[
1-\frac{1}{\left(  1+\Omega_{0}\right)  }\left(  \frac{\Omega_{m0}}{a}%
-\frac{\Omega_{v\gamma0}}{a^{2}}\right)  \right]  . \tag{83}%
\end{equation}
Because of $(80)$, $V-space$ will expand with a deceleration in the initial period.

\subsection{Expansion of space}

\subsubsection{To determine $a\left(  t\right)  $}

From $(83)$ we have
\begin{equation}
t=t_{0}-\frac{1}{H_{0}\sqrt{1+\Omega_{0}}}\left\{  \sqrt{1-M+\Gamma}%
-\sqrt{a^{2}-Ma+\Gamma}+\frac{M}{2}\ln\frac{2-M+2\sqrt{1-M+\Gamma}%
}{2a-M+2\sqrt{a^{2}-Ma+\Gamma}}\right\}  , \tag{84}%
\end{equation}
where $M=\Omega_{m0}/\left(  1+\Omega_{0}\right)  ,$ $\Gamma=\Omega_{v\gamma
0}/\left(  1+\Omega_{0}\right)  .$

As mentioned above, we suppose that $T-space$ transforms $V-space$, i.e.,
$\left\langle \omega_{v}\right\rangle =0\longrightarrow\varpi_{v0}\neq0$ and
$\left\langle \omega_{s}\right\rangle =0\longrightarrow0.$ Hence the Higgs
potential energy $V_{0}$ must first transform into $\rho_{v}^{\prime}=\alpha
V_{0}.$ On the other hand, because $m\left(  \omega_{v}\right)  \sim0$ and
$m\left(  \omega_{s}\right)  \sim0$ when $T_{v}\sim T_{s}\sim T_{\max},$
$v-Higgs$ particles $\omega_{v}$ can transform into $s-Higgs$ particles
$\omega_{s}$ by the coupling $(13)$, i.e., a part of $V_{0}$ finally transform
into $s-Higgs$ particles $\omega_{s}.$ Let $\rho_{s}^{\prime}=(1-\alpha
)V_{0}\equiv\beta V.$ It is necessary $\alpha>\beta$.

$V-matter$ can exist in many forms, e.g., $v-visible$ matter and $v-dark$
matter, $v-particles$ with their large masses and $v-particles$ without mass
or very small mass (e.g., neutrino). It is possible there are some $v-dark$
particles without mass or with very small mass. Here $\rho_{v\gamma\text{ }}%
$contains all (visible and dark) $v-particles$ whose masses are zero or very
small. Because we cannot completely understand dark matter now, we cannot
determine $\rho_{v\gamma}.$ In contrast with $v-matter$, $s-matter$ exists in
only one form, i.e., $SU(5)$ colour single states whose masses are not zero.
Consequently, although $\rho_{v}=\rho_{vm}+\rho_{v\gamma}>\rho_{s}=\rho_{sm},$
it is possible $\rho_{sm}>\rho_{vm}$ and $\rho_{sm}>\rho_{s}$ when $R$ is big enough.

Taking $\Omega_{v\gamma0}=0.001,$ $\Omega_{m0}=0.3\Omega_{v\gamma0}%
+2\sqrt{\Omega_{v\gamma0}},$ $H_{0}^{-1}=9.7776\times10^{9}h^{-1}yr^{\left[
12\right]  }$ and $h=0.8,$ we get the $a\left(  t\right)  $ which showed by
the curve $B$ in the figure 1 and describes evolution of the cosmos from
$14\times10^{9}yr$ ago to now; Taking $\Omega_{v\gamma0}=0.05,$ $\Omega
_{m0}=2\sqrt{\Omega_{v\gamma0}}$ we get the $a\left(  t\right)  $ which showed
by the curve $A$ in the figure 1 and describes evolution of the cosmos from
$13.7\times10^{9}yr$ ago to now. Provided $\Omega_{m0}\leqslant2\Omega
_{v\gamma0}+2\sqrt{\Omega_{v\gamma0}}$ (the condition $\overset{\cdot}{a}%
^{2}>0),$ we can get a curve of $a\left(  t\right)  $ which describes
evolution of the cosmological scale.

From the two curves we see that the cosmos must undergo a period in which the
cosmos expands with a deceleration and the present period in which the cosmos
expands with a acceleration.

Ignoring $\Omega_{v\gamma0},$ $\Omega_{m0}\longrightarrow-\Omega_{m0}$ and
taking $a\sim0$, we can reduce $(84)$ to the corresponding formula$^{\left[
12\right]  }.$

\subsubsection{The relation between redshift and optical distance}

Considering $k=-1,$ from $(40)$ and $(81)$ we have%

\begin{equation}%
%TCIMACRO{\dint \nolimits_{a}^{1}}%
%BeginExpansion
{\displaystyle\int\nolimits_{a}^{1}}
%EndExpansion
\frac{cda}{R\overset{\cdot}{a}}=-%
%TCIMACRO{\dint \nolimits_{r}^{0}}%
%BeginExpansion
{\displaystyle\int\nolimits_{r}^{0}}
%EndExpansion
\frac{dr}{\sqrt{1+r^{2}}}, \tag{85}%
\end{equation}%
\begin{align}
H_{0}d_{L}  &  =H_{0}R_{0}r(1+z_{d})=\frac{2c}{\left(  \Omega_{m0}%
-2\Omega_{v\gamma0}\right)  ^{2}-4\Omega_{v\gamma0}}\cdot\nonumber\\
&  \left\{  2\left(  1+\Omega_{0}\right)  -\left(  1+z_{d}\right)  \Omega
_{m0}-\left[  2\left(  1+\Omega_{0}\right)  -\Omega_{m0}\right]
\sqrt{1-\left(  \Omega_{m0}-2\Omega_{v\gamma0}\right)  z_{d}+\Omega_{v\gamma
0}^{2}z_{d}^{2}}\right\}  , \tag{86}%
\end{align}
where $z_{d}=\left(  1/a\right)  -1$ is the redshift caused by $R$ increasing.
Provided $\Omega_{m0}\longrightarrow-\Omega_{m0},$ $(86)$ is consistent with
the corresponding formula$^{\left[  12\right]  }.$ Ignoring $\Omega_{v\gamma
0},$ $\Omega_{m0}\longrightarrow-\Omega_{m0}$ we reduce $(86)$ to%
\begin{equation}
H_{0}d_{L}=\frac{2c}{\Omega_{m0}^{2}}\left\{  2+\Omega_{m0}\left(
1-z_{d}\right)  -\left[  2+\Omega_{m0}\right]  \sqrt{1-\Omega_{m0}z_{d}%
}\right\}  . \tag{87a}%
\end{equation}
Approximating to $\Omega_{m0}^{1}$ and $z_{d}^{2}$, we obtain
\begin{equation}
H_{0}d_{L}=z_{d}+\frac{1}{2}z_{d}^{2}\left(  1+\frac{1}{2}\Omega_{m0}\right)
. \tag{87b}%
\end{equation}

Taking $\Omega_{v\gamma0}=0.001,$ $\Omega_{m0}=0.3\Omega_{v\gamma0}%
+2\sqrt{\Omega_{v\gamma0}}$ and $H_{0}^{-1}=9.7776\times10^{9}h^{-1}%
yr^{\left[  12\right]  }$ and $h=0.8,$ we get the $d_{L}-z_{d}$ relation which
showed by the curve $A$ in the figure 2; Taking $\Omega_{v\gamma0}=0.05,$
$\Omega_{m0}=2\sqrt{\Omega_{v\gamma0}}$ we get the $d_{L}-z_{d}$ relation
which showed by the curve $B$ in the figure 2.

From $(84)$, $(86)$, figure 1 and 2 we can choose suitable parameters to get
the curves describing true evolution of the cosmos.

\section{Dark energy and visible energy transform each other and action of
dark energy on a galaxy}

\subsection{Dark energy and visible energy transform each other}

As mentioned above, when temperature $T_{s}$ $\sim T_{\max}$ in $S-space,$
$\langle\omega_{s}\rangle\longrightarrow0$ and $\langle\omega_{v}\rangle=0$,
the masses of Higgs bosons tend to zero. Consequently $\Omega_{s},$ $\Phi_{s}$
and $\chi_{s}$ can be enormously produced and easily transform into
$\Omega_{v},$ $\Phi_{v}$ and $\chi_{v}$ by the couplings in $(13),$ and there
is thermal equilibrium between the $v-particles$ (the dark energy in
$S-space$) and the $s-particles$ when $T_{s}\sim T_{v}\sim T_{\max}$.

When the temperature $T_{s}$ is low, there is no thermal equilibrium between
the $v-particles$ and the $s-particles$. But the repulsive potential energy
between $v-matter$ and the $s-matter$ can transform into the energy of the
dark energy.

The repulsion potential energy is determined by the distributing mode of
$s-matter$ and $v-matter.$ In $V-space$ $v-particles$ with their masses can
form celestial bodies. There is no interaction except gravity among $S-SU(5)$
color single states, hence the color single states cannot form any dumpling
and must relaxedly distribute in $V-space.$ Consequently, the huge repulsion
potential energy between the $V-celestial$ bodies and the $S-SU(5)$ color
single states must chiefly transform into the kinetic energy of $S-SU(5)$
color single states, i.e., the dark energy in $V-space$. In fact, we can prove
in flat space that when space expands $K$ times, i.e., $R\longrightarrow KR,$
the repulsive-potential energy density $V_{r}$ becomes $V_{r}/K$ and%
\begin{equation}
\bigtriangleup V_{r}=\left(  1-1/K\right)  V_{r}. \tag{88}%
\end{equation}
To consider a system in flat space which is composed of a $v-body$ with its
mass $M$ and a $s-colour$ single state with its mass $m.$ It is easily to get
the rate $\bigtriangleup E_{m}/\bigtriangleup E_{M}$ for static $M$ and $m$ at
the initial moment..
\begin{equation}
\frac{\bigtriangleup E_{m}}{\bigtriangleup E_{M}}=\frac{2M+\bigtriangleup
V_{r}}{2m+\bigtriangleup V_{r}} \tag{89}%
\end{equation}
Because $M\gg m$, $\bigtriangleup E_{m}>\bigtriangleup E_{M}.$ Thus, although
$\rho_{v}>\rho_{s}$ after inflating$,$ after $V-space$ expands with a
deceleration for some time, it is also possible $\rho_{v}<\rho_{s}.$ This
implies that $\rho_{v}-\rho_{s}=\rho$ is changeable, can change from $\rho>0$
to $\rho<0$ as expanding of $V-space$.

Space expansion is not the necessary condition to transform repulsive
potential energy into kinetic energy. Let $\overset{\cdot}{R}_{V}=0$ or $<0.$
Because $\rho_{v}$ is not absolutely even, some $v-matter$ can gather to a
region and forms a galaxy by its gravity. Thus $s-matter$ which is initially
in the region must be repelled away the region and increases its kinetic
energy, and the repulsive potential energy chiefly transforms into the kinetic
energy of the $s-matter$ in the case as well. Thus, although $\rho_{v}%
\gtrsim\rho_{s}$ when $\overset{\cdot}{R}_{V}=0$ or $<0,$ it is also possible
that $\rho_{v}\gtrsim\rho_{s}$ evolves to $\rho_{v}<\rho_{s}$ since $v-matter$
gathers and forms galaxies. Because there is no interaction except the
repulsion, a $v-observer$ will regard that energy seems to come into being or
annihilate at the highest temperature $T_{\max}$.

\subsection{Steadiness of galaxies}

According to the conventional theory, when space expands with an acceleration,
a galaxy will suffer a force $F$ which causes the volume\ of the galaxy to
increase, and if the acceleration is big enough, the galaxy will break apart.
But according to the present theory, this cannot occur. As mentioned above,
expanding of $V-space$ is because there is $s-matter,$ more precisely, is
because $\rho_{v}-\rho_{s}=\rho<0,$ and expanding acceleration is
$\overset{\cdot\cdot}{R}=\eta\rho R_{0}^{3}/2R^{2}.$ The force $F$ caused by
accelerative expanding is directly proportional to $\overset{\cdot\cdot}{R}.$
On the other hand, because there is repulsion between $s-matter$ and
$v-matter,$ the $v-galaxy$ also suffers the pressure $P$\ coming from
$s-matter$ surrounding it. If we regard $s-matter$ and $v-matter$ as
relatively static, i.e., they move only with expanding of $V-space,$ the
pressure $P$ suffered by the $v-galaxy$ will directly proportional to
$\eta\rho R_{0}^{3}/2R^{2}$ as well. Thus $P/F$ is irrelative to $R.$ Because
the directions of the two forces $P$ and $F$ are opposite, the resultant of
$P$ and $F$ on the $v-galaxy$ is direct proportion to $\eta\rho R_{0}%
^{3}/2R^{2}$, and can be lesser or approach to zero as $R$ increases. Thus,
although $V-space$ expands with an acceleration, $v-galaxies$ can still be steady.

The pressure $P$\ on a visible $v-galaxy$ coming from $s-matter$ is similar to
the gravity of $v-dark$ matter inside the visible $v-galaxy.$

\section{Repulsive redshift}

As mentioned above, in $V-space$ $v-matter$ can form galaxies, $s-matter$
cannot form dumplings and can only relaxedly distribute in $V-space.$ A
$v-particle$ static relatively to $s-matter$ suffers the resultant of
repulsive forces coming from $s-matter$ to be zero. But when a $v-particle$,
e.g. a $v-photon$, moves relatively to $s-matter$, it suffers the resultant of
repulsive forces coming from $s-matter$ not to be zero. This is because the
density of $s-energy$ in the front of the $v-photon$ seems higher than that in
the rear of the $v-photon$ Thus the $v-photon$ will suffer a repulsive force
in the direction opposing to its velocity. Analogously to gravitational shift,
the $v-photon$ will have a repulsive redshift in $V-space$. The repulsive
redshift of the photon should be directly proportional to the density
$\rho_{s},$ the energy of the photon and the moving distance $l=ct=t$ of the
photon (where $c=1)$. Thus we have%

\begin{equation}
d\nu=-f\left(  \rho_{s}-\rho_{v}\right)  \nu dl=-f\left(  \rho_{s}\left(
t\right)  -\rho_{s}\left(  t\right)  \right)  \nu\left(  t\right)
dt\equiv-f\rho\left(  t\right)  \nu\left(  t\right)  dt, \tag{90}%
\end{equation}
where .$f\sim\eta$ is a constant$.$\ 

\subsection{The case $k_{V}=-1$ and $\rho\equiv\rho_{vm}-\rho_{sm}<0$}

Considering $\rho_{sm}\left(  t\right)  =\rho_{sm0}/a^{3}\left(  t\right)  ,$
$\rho_{vm}\left(  t\right)  =\rho_{vm0}/a^{3}\left(  t\right)  $ so that
$\rho\left(  t\right)  =\rho_{m0}/a^{3}\left(  t\right)  ,$ ignoring
$\rho_{v\gamma}$ and taking $c=1$, from $\left(  81\right)  $ and $\left(
90\right)  ,$ we have
\begin{equation}%
%TCIMACRO{\dint \limits_{\nu}^{\nu_{0}}}%
%BeginExpansion
{\displaystyle\int\limits_{\nu}^{\nu_{0}}}
%EndExpansion
\frac{d\nu}{\nu}=-f\rho_{0}%
%TCIMACRO{\dint \limits_{a}^{1}}%
%BeginExpansion
{\displaystyle\int\limits_{a}^{1}}
%EndExpansion
\frac{da}{a^{3}\overset{\cdot}{a}}=-f\rho_{0}R_{0}%
%TCIMACRO{\dint \limits_{a}^{1}}%
%BeginExpansion
{\displaystyle\int\limits_{a}^{1}}
%EndExpansion
\frac{da}{a^{\prime2}\sqrt{a^{\prime2}-Q_{0}a^{\prime}}}, \tag{91}%
\end{equation}%
\begin{align}
\frac{\nu}{\nu_{0}}  &  =\exp\frac{2}{3}\frac{A}{Q_{0}}\left\{  \left(
1+\frac{2}{Q_{0}}\right)  \sqrt{1-Q_{0}}-\left(  1+\frac{2}{Q_{0}}%
+z_{d}\right)  \sqrt{1-\left(  1+z_{d}\right)  Q_{0}}\right\}  ,\tag{92}\\
A  &  \equiv\frac{f}{\eta}\frac{Q_{0}}{R_{0}}=\frac{f}{\eta}H_{0}^{2}%
R_{0}\Omega_{s0}=f\left(  \rho_{s0}-\rho_{v0}\right)  R_{0},\text{
\ \ \ }Q_{0}<1 \tag{93}%
\end{align}
Thus the red shift caused by repulsion is%
\begin{equation}
z_{r}=\frac{\nu}{\nu_{0}}-1. \tag{94}%
\end{equation}
$A$ is very small, e.g., for $Q_{0}=1/2$, $A=f/2\eta R_{0}.$ Approximately to
$A,$ we have%

\begin{equation}
z_{r}\simeq A\left(  z_{d}+\frac{1}{2}z_{d}^{2}\right)  , \tag{95}%
\end{equation}
A red shift $z$ observed by us satisfies$^{\left[  12\right]  }$,
\begin{equation}
\left(  1+z\right)  =\left(  1+z_{r}\right)  \left(  1+z_{d}\right)  .
\tag{96}%
\end{equation}
We should represent the relation between a redshift and distance by total
red-shift. From $\left(  94\right)  -\left(  96\right)  $ we have%
\begin{equation}
z_{d}=z\frac{z_{d}}{z}\equiv zq^{-1},\text{ \ \ \ }q=1+z_{r}/z_{d}+z_{r}.
\tag{97}%
\end{equation}
Approximately to $A$ we have
\begin{equation}
q\simeq1+A\left(  1+\frac{3}{2}z+\frac{1}{2}z^{2}\right)  ,. \tag{98}%
\end{equation}
Considering $z_{d}=zq^{-1},$ approximately to $A$ and $z^{2}$ when $z<1,$ from
$\left(  87b\right)  $ we have%
\begin{equation}
\left(  1+A\right)  H_{0}d_{L}\equiv H_{0eff}d_{L}\simeq z+\frac{1}{2}%
z^{2}\left\{  \left(  1+\frac{1}{2}\Omega_{mo}\right)  -A\left(  \frac{5}%
{2}+\frac{1}{2}\Omega_{mo}\right)  \right\}  . \tag{99}%
\end{equation}
where $H_{0eff}=\left(  1+A\right)  H_{0}$ is the effective Hubble constant.

From $\left(  1+z_{d}\right)  ^{-1}=a\left(  t\right)  $ and $\left(
95\right)  -\left(  96\right)  $ we obtain%
\begin{align}
\widetilde{a}\left(  t\right)   &  =\frac{\widetilde{R}\left(  t\right)
}{R_{0}}\equiv\left(  1+z\right)  ^{-1}=\left(  1+z_{d}\right)  ^{-1}\left(
1+z_{r}\right)  ^{-1}\nonumber\\
&  \simeq\left(  1+z_{d}\right)  ^{-1}\left[  1+A\left(  z_{d}+\frac{1}%
{2}z_{d}^{2}\right)  \right]  ^{-1}=a\left(  t\right)  \left[  1-\frac{1}%
{2}A\left(  \frac{1}{a^{2}}-1\right)  \right]  . \tag{100}%
\end{align}
where $\widetilde{R}\left(  t\right)  $ is not the true scalar curvature, but
is an apparent scalar curvature determined by the total red shift $z.$ From
$\left(  100\right)  $ we can determine $\widetilde{a}\left(  t\right)  $
after $a\left(  t\right)  $ is determined$.$ Because $A$ is very small, the
correction coming from $z_{r}$ is very small.

\section{Structure of the universe}

\subsection{The universe is composed of infinite cosmic islands}

If the whole universe is the $s-world$ or the $v-world$, the analysis above is
correct. But according to the present model, there possibly is new structure
of the universe.

As mentioned above, $v-matter$ and $s-matter$ are symmetric and mutually
repulsive, $v-matter$ in $v-space$ can form a $v-world$ and $s-matter$ in
$s-space$ can form a $s-world$. From this we present a new cosmic model to be
that the universe is composed of infinite cosmic islands and every cosmos
island is just a $s-world$ in $s-space$ or a $v-world$ in $v-space.$ Here both
$v-space$ and $s-space$ are finite.

According to the model we easily see that there must be only $v-cosmic$
islands neighboring a $s-cosmic$ island. This is because if two $s-cosmic$
islands are neighboring, they will combine to form one new larger $s-cosmic$
island by their gravitation. It is necessary that the distance between two
$v-cosmic$ islands must be very huge. Because of the same reason there are
only $s-cosmic$ islands around a $v-cosmic$ island. Because there is no
interaction except the repulsion between $v-matter$ and $s-matter$, there is
no interaction except the repulsion between a $s-cosmic$ island and a
$v-cosmic$ islands as well. Because of the following reasons, we in a
$v-cosmic$ island cannot observe anything of the $s-cosmic$ islands around us.

All Higgs particles whose masses are not zero will fast decay into fermions
and gauge bosons. The masses of $V-fermions$ and $v-gauge$ bosons in a
$s-cosmic$ island are all zero and must form $v-SU(5)$ colour single states.
The masses of $V-fermions$ and $v-gauge$ bosons in a $v-cosmic$ island are not
zero (the masses of $v-photons$ and $v-glutons$ in $v-cosmic$ island are
zero). Hence a $v-SU(5)$ colour single states in a $s-cosmic$ island cannot
come into the $v-cosmic$ island in which $v-SU(5)$ has broken to
$v-SU(3)\times U(1).$

A $s-particle$ in a $s-cosmic$ island must be $s-SU(5)$ non-colour single
state. If a $s-particle$ came into the $v-cosmic$ island, it would still be
non-colour single state and would have very big mass. Hence a $s-particle$ in
a $s-cosmic$ island cannot come into the $v-cosmic$ island. Because of the
same reason, no particle can fly away the $v-cosmic$ island.

As a consequence a $v-observer$ in the $v-cosmic$ island can regard the
$v-cosmic$ island as the whole world.

A $v-cosmic$ islands and a $s-cosmic$ island can influence each other by the
Higgs potential in their boundary. For example, $\rho_{s}\sim0$ and $\rho
_{v}\sim0$ in a $v-cosmic$ island, the expectation values of the $v-Higgs$
field in the island will be assimilated by the $s-cosmic$ islands neighboring
to it. Hence the $v-cosmic$ island will disappear.

Thus we can approximately regard a cosmic island as the whole cosmos. It is
possible that some cosmic islands are forming or expanding, and the other
cosmic islands are contracting or disappear. It is also possible that some
cosmic islands anew combine to form a new larger cosmic island due to their
motion or expansion. The cosmic island in which we exist is expanding.

Thus, according to the present model the cosmos as a whole is infinite and its
properties are always unchanging, and there is no starting point or end of time.

\subsection{\textbf{Redshifts and characters of quasars}}

Hydrogen spectrum is
\begin{equation}
\omega_{nk}=(E_{n}-E_{k})/\hbar=-\frac{\mu e^{4}}{2\hbar^{3}}(\frac{1}{n^{2}%
}-\frac{1}{k^{2}}),\text{ \ \ }\mu=\frac{mM}{m+M}, \tag{101}%
\end{equation}
where $m$ is the mass of an electron, and $M$ is the mass of a proton.
According the unified model, $m\propto\upsilon_{e}$, the mass of a quark
$m_{q}\propto\upsilon_{q},$ where $\upsilon_{e}$ and $\upsilon_{q}$ are the
expectation values of the Higgs fields coupling with the electron and the
quark, respectively. Consequently the mass of a proton is approximately
$M\propto\upsilon_{q}.$

Because the cosmic islands neighboring with a $s-cosmic$ island in which
$\varpi_{v}=0$ and $\varpi_{s}=$ $\varpi_{s0}$ must be $v-cosmic$ islands in
which $\varpi_{v}=\varpi_{v0}$ and $\varpi_{s}=$ $0,$ it is necessary there is
the transitional region in which $\varpi_{v}=0\longrightarrow\varpi_{v0}$ and
$\varpi_{s}=$ $\varpi_{s0}\longrightarrow0$ in the boundary of the $s-cosmic$
island, i.e., $0\leq\mid\varpi_{s}\mid\leq\mid\varpi_{s0}\mid$ and $0\leq
\mid\varpi_{v}\mid\leq\mid\varpi_{v0}\mid$ in the transitional region.
Consequently the masses $m^{\prime}$, $M^{\prime}$ and $\mu^{\prime}$ in the
transitional regions must be less than the conventional masses. From $(129)$
we see the characteristic frequency $\omega_{nk}^{\prime}$ must be less than
$\omega_{nk}$,
\begin{equation}
\bigtriangleup\omega_{nk}=\omega_{nk}-\omega_{nk}^{\prime}=-\frac{(\mu
-\mu^{\prime})e^{4}}{2\hbar^{3}}(\frac{1}{n^{2}}-\frac{1}{k^{2}}). \tag{102}%
\end{equation}
The sort of red-shifts is called $mass$ $redshift$. The mass redshift is
essentially different from the red-shift mentioned above. Thus, the photons
originating from the star in the transitional region must have larger
red-shift than that determined by the Hubble formula at the same distance.
Thereby we guess that some quasars are just the galaxies in the transitional
region of our cosmos island.

An ordinary $s-galaxy$ and a $s-quasar$ can be neighboring, because a
transitional region in which $\mid\varpi_{s}\mid\leq\mid\varpi_{s0}\mid$ and
$\mid\varpi_{v}\mid\leq\mid\varpi_{v0}\mid$ must be neighboring to an ordinary
region in which $\varpi_{s}=\varpi_{s0}$ and $\varpi_{v}=0$. The difference of
the redshifts of the neighboring ordinary $s-galaxy$ and the $s-quasar$ is
very large.

If an atom or a star run from the ordinary region to the transitional region
with a high velocity, what will occur? the atom or the star will explode
because of mass change of the particles composing the atom or the star. We
will discuss the problem in other paper.

\section{New predictions}

$\mathbf{1}$\textbf{. It is possible that some huge cavity is equivalent to a
huge concave lens. }

According the present model, it is possible that some huge cavities in
$V-space$ are not empty, but there are $\rho_{s}^{\prime}>\rho_{s}$ and
$\rho_{v}^{\prime}\ll\rho_{v}$ in it so that $\rho^{\prime}\equiv\rho
_{s}^{\prime}-\rho_{v}^{\prime}>\rho=\rho_{s}-\rho_{v},$ here $\rho
_{s}^{\prime}$ and $\rho_{v}^{\prime}$ denote the densities in the cavity, and
$\rho_{s}$ and $\rho_{v}$ denote the average densities. Because there are the
repulsion between $s-matter$ and $v-matter$ and the gravity among $s-matter,$
a huge cavity can form. The characters of such a huge cavity are as follows.

$\mathbf{A}$. Because there is no interaction similar to the electromagnetic
interaction, except the gravity, among the $s-SU(5)$ colour single states and
the masses of the $s-SU(5)$ colour single states are very small, a cavity in
$V-space$ must be very large.

$\mathbf{B}$. Because $s-matter$ and $v-matter$ are mutually repulsive, when
$v-photons$ pass through such a huge cavity, the $v-photons$ must suffer
repulsion and consequently are scattered as they pass through a huge concave
lens, the galaxies behind the huge cavity seem to be darker and more remote,
and $v-photons$ pass through such a huge cavity will have larger redshift.

$\mathbf{C}$. Because there is the repulsion from $s-matter$ to the background
radiation, the density of the background radiation inside the huge cavity must
very little. Hence the temperature of the huge cavity must very low. It is
seen that the present model can well explain the characters of some huge cavities.

This is a decisive prediction which distinguishes the present model from other models.

$\mathbf{2.}$\textbf{ The universal total energy observed by us seems not to
be conservational.}

The universal total energy, of course, is conservational in fact. Because the
energies $s-matter$ and $v-matter$ can transform from one into another at the
highest temperature and a $v-observer$ in $V-space$ cannot detect $s-matter$
except by gravity, the universal total energy observed by the $v-observer$
seems not to be conservational. As a consequence, the cause of space expanding
with an acceleration seems to be because of mass loses or the gravitation decreases.

$\mathbf{3}$\textbf{. The gravitation between two galaxies distant enough will
lesser than that predicted by the conventional theory. }

There must be $s-matter$ relaxedly distributing between two $v-galaxies$
distant enough, hence the gravitation between them must lesser than that
predicted by the conventional theory. But when the distance between the two
$v-galaxies$ is small, the gravitation between them is not influenced by
$s-matter$ because when $\rho_{v}$ is big, $\rho_{s}$ must be small.

$\mathbf{4}.$ \textbf{Transformation of a black hole with its mass big enough}

Let there is a $v-brack$ hole with its mass big enough in $V-space.$ If its
mass is so big that its temperature to be $\sim T_{\max},$ the expectation
values of the Higgs fields inside the $v-brack$ hole will change, $\varpi
_{v}=\varpi_{v0}\longrightarrow0$ and $\varpi_{s}=0$, so that the space will
expand and finally $\varpi_{v0}=0$ and $\varpi_{s}=0\longrightarrow\varpi
_{s0}$. Consequently, the $v-brack$ hole transforms into a $s-cosmic$ island
which is different from $v-huge$ cavities in $V-space$. A $v-observer$ can
observe it by its repulsive effects, but cannot get other messages.

$\mathbf{5}$\textbf{. A sort of huge white holes}

Let there be a $s-cosmic$ island neighboring on a $v-cosmic$ island. It is
possible that the $s-cosmic$ island transforms into a $v-cosmic$ island after
it contraction. In this case, a $v-observer$ in the $v-cosmic$ island must
observe a very huge white huge.

There are other predictions. We will discuss them in following paper.

\section{Conclusions}

The new hypotheses are proposed that there are $s-matter$ and $v-matter$ which
are symmetric and mutually repulsive, there are $S-space$ and $V-space$ whose
essential difference is that their expectation values of the Higgs fields are
different. In $S-space$ the $S-SU(5)$ symmetry is broken into $S-SU(3)\times
U\left(  1\right)  $ for $s-particles$ and $V-SU(5)$ symmetry still strictly
holds for $v-particles.$ As a consequence $s-particles$ get their masses
determined by the $SU(5)$ $GUT$ and form the $s-world$, and $v-particles$
without mass form $V-SU(5)$ color-single states which are identified with dark
energy in $S-space.$ Evolving process of the universe is as follows. Firstly
$S-space$ contracts due to $\rho_{s}-\rho_{v}$ large enough and causes the
temperature $T_{s}$ to rise. When $T_{s}\longrightarrow T_{\max}$, both
$V-SU(5)$ and $S-SU(5)$ symmetries strictly hold and therefore the masses of
all particles are zero. Secondly inflation of $V-space$ occurs. After the
inflation the phase transiting of the vacuum occurs. As a consequence
$S-space$ transforms into $V-space$ and the $s-world$ transforms into the
$v-world.$ The results above still hold water when $V\rightleftarrows S$ and
$v\rightleftarrows s.$ The following results are obtained. There is no
spacetime singularity. There is the highest temperature and the highest energy
density in the universe. The results of the Guth's inflationary scenario are
obtained. $Space$ seems to expand with a deceleration in early period and with
an acceleration in later period. A formula which well describes the luminous
distance and redshift is obtained. The universe is composed of infinite cosmic
islands which are $s-worlds$ or $v-worlds$. The new predictions of the present
model are as follows. Some huge cavity is not empty and is equivalent to a
huge concave lens. The present model can well explains the given characters of
some huge cavities. The universal total energy is conservational, but the
total energy observed by us seems not to be conservational. The gravitation
between two galaxies distant enough will lesser than that predicted by the
conventional theory. A possible explanation for the big redshifts of
quasi-stellar objects is presented. The huge redshifts of quasars are the mass
redshifts. It is possible that a $v-brack$ hole with its mass big enough can
transform into $s-huge$ white hole which cannot be observed by a $v-observer$
except by its repulsion effects, and the $s-cosmic$ island neighboring on a
$v-cosmic$ island transforms into a $v-cosmic$ island after it contraction. In
this case, a $v-observer$ in the $v-cosmic$ island must observe a very huge
$v-huge$ white holes which are different from that predicted by conventional theory.

Before inflating occurs, the $s-world$ is in thermal equilibrating state. If
there is no $v-matter,$ because of contraction by gravitation, the $s-world$
would become a thermal equilibrating singular point, i.e., the $s-world$ would
be in the hot death state. As seen, it is necessary that there are both
$s-matter$ and $v-matter$ and both $S-space$ and $V-space$.

\section{Discussion}

\subsection{$k$ in the Robertson-Walker metric is regarded as a function of
density $\rho$}

For spatially homogeneous and isotropic 3-dimensional space, the Riemann
curvature tensor describing the 3-dimensional geometry at any moment
satisfies$^{\left[  13\right]  .\text{ }}.$%
\begin{equation}
^{(3)}R_{mnsk}=K(t)(\delta_{m}^{s}\delta_{n}^{k}-\delta_{m}^{k}\delta_{n}%
^{s}), \tag{D1}%
\end{equation}
The Robertson-Walker metric $\left(  40\right)  $ satisfies $\left(
D1\right)  $ and can equivalently be rewritten as$^{\left[  13\right]  }$%
\begin{equation}
ds^{2}=dt^{2}-[\frac{dr^{2}}{1\pm r^{2}/a\left(  t\right)  ^{2}}+r^{2}%
(d\theta^{2}+\sin^{2}\theta d\varphi^{2}]. \tag{D2}%
\end{equation}
We consider it is possible that $k$ can change from $\sim1$ to $0$ and further
to $\sim-1$ which cannot attribute to change of $R(t).$ In the conventional
theory, because there is one sort of matter so that it is necessary $\rho>0$,
it is unnecessary to consider $k$ to be changeable. But according to the
present model, $\rho>0$, $\rho=0$ and $\rho<0$ are all possible$,$ here
$\rho=\rho_{v}-\rho_{s},$ hence it is necessary to consider $k$ to be
changeable. Thus we rewrite $\left(  40\right)  $ as
\begin{equation}
ds^{2}=dt^{2}-R^{\prime2}(t)[\frac{dr^{2}}{1-r^{2}/A\left(  \rho\right)
}+r^{2}(d\theta^{2}+\sin^{2}\theta d\varphi^{2}] \tag{D2}%
\end{equation}
$(D2)$ is invariant under transformation of $A\left(  \rho\right)
\longrightarrow A\left(  \rho\right)  /\left\vert A\left(  \rho\right)
\right\vert $, $r\longrightarrow r/\sqrt{\left\vert A\left(  \rho\right)
\right\vert }$ and $R\longrightarrow\sqrt{\left\vert A\left(  \rho\right)
\right\vert }R$. If $A\left(  \rho\right)  $ is only a positive or negative
real function or infinite, $(D2)$ is equivalent to $(40)$. But we suppose
$A\left(  \rho\right)  $ is a real function of the density in comoving
coordinate system and can change from $A\left(  \rho\right)  >0$ to $\infty$
and further to $A\left(  \rho\right)  <0.$ Thus $(D2)$ is no longer equivalent
to $\left(  40\right)  .$

From $(51)$ or $(53)$ we see when $k=1,$ $\rho>0$ is necessary, because there
is no real solution in this case corresponding to $\rho\leq0.$ When $k=0,$
$\rho\geq0$ is necessary, because there is no real solution corresponding to
$\rho<0$ in this case$.$ When $k=-1,$ there are real solutions for $\rho\geq0$
or $\rho<0.$ Thus, we suppose, e.g. in $V-space,$ that $k$ alters as the
density in comoving coordinate system%
\begin{align}
k\left(  \rho\right)   &  \equiv A^{-1}\left(  \rho\right)  \equiv
\frac{V+\left(  \rho_{m}+\rho_{\gamma}\right)  R^{3}/R_{1}^{3}}{V_{1}+\left(
\rho_{m1}+\rho_{\gamma1}\right)  },\text{ \ \ for }V_{1}+\left(  \rho
_{m1}+\rho_{\gamma1}\right)  >0,\tag{D3a}\\
k\left(  \rho\right)   &  \equiv A^{-1}\left(  \rho\right)  \equiv
-\frac{V+\left(  \rho_{m}+\rho_{\gamma}\right)  R^{3}/R_{1}^{3}}{V_{1}+\left(
\rho_{m1}+\rho_{\gamma1}\right)  },\text{ \ for }V_{1}+\left(  \rho_{m1}%
+\rho_{\gamma1}\right)  <0, \tag{D3b}%
\end{align}
where $\rho_{m}\equiv\rho_{vm}-\rho_{sm},$ $\rho_{\gamma}\equiv\rho_{v\gamma
}-\rho_{s\gamma},\rho\left(  t\right)  =\rho_{m}\left(  t\right)
+\rho_{\gamma}\left(  t\right)  =\rho\left(  t,R\right)  $, $\rho_{1}%
\equiv\rho\left(  t_{1}\right)  ,$ $V$ is the Higgs potential energy density,
and $\gamma$ denotes all massless particles. In general, although $R$ does not
change, $\rho_{vm}-\rho_{sm}$ can also alter since the repulsive potential
energy chiefly transforms kinetic energy of $s-colour$ single states when
$v-celestial$ bodies form. On the other hand, change of $R$ must cause $\rho$
to change. Hence $\rho=\rho\left(  t,R\left(  t\right)  \right)  .$ When
$T=T_{\max}$, $V=V_{0},$ $\rho\sim0$ and
\begin{equation}
k\left(  \rho\right)  =V_{0}/\left(  V_{1}+\rho_{m1}+\rho_{\gamma1}\right)  .
\tag{D4a}%
\end{equation}
After reheating, when temperature is low so that $V\sim V_{1}\sim0,$ we have%
\begin{equation}
k\left(  \rho\right)  =\frac{\left(  \rho_{m}+\rho_{\gamma}\right)
R^{3}/R_{1}^{3}}{\left(  \rho_{m1}+\rho_{\gamma1}\right)  }\text{ \ for }%
\rho_{1}>0,\text{ \ }k\left(  \rho\right)  =-\frac{\left(  \rho_{m}%
+\rho_{\gamma}\right)  R^{3}/R_{1}^{3}}{\left(  \rho_{m1}+\rho_{\gamma
1}\right)  }\text{ \ \ for }\rho_{1}<0. \tag{D4b}%
\end{equation}
When $\rho\left(  t,R\right)  =\rho\left(  R\right)  ,$ $\rho_{m}R^{3}%
=\rho_{m1}R_{1}^{3}$ and $\rho_{\gamma}R^{4}=\rho_{\gamma1}R_{1}^{4},$ we have
$k\left(  \rho\left(  t,R\right)  \right)  =k\left(  \rho\left(  R\right)
\right)  ,$
\begin{equation}
k\left(  \rho\right)  =\frac{\rho_{m1}+\rho_{\gamma1}\left(  R_{1}/R\right)
}{\rho_{m1}+\rho_{\gamma1}},\text{ \ for }\rho_{1}>0;\text{ \ }k\left(
\rho\right)  =-\frac{\rho_{m1}+\rho_{\gamma1}\left(  R_{1}/R\right)  }%
{\rho_{m1}+\rho_{\gamma1}}\text{ \ for }\rho_{1}<0 \tag{D4c}%
\end{equation}
Let $k\left(  \rho\left(  R^{\prime}\right)  \right)  =0$ for $\rho_{1}>0,$ we
have $k\left(  \rho\left(  R>R^{\prime}\right)  \right)  <0$ and $k\left(
\rho\left(  R<R^{\prime}\right)  \right)  >0$; and for $\rho_{1}<0,$ we have
$k\left(  \rho\left(  R>R^{\prime}\right)  \right)  <0$ and $k\left(
\rho\left(  R<R^{\prime}\right)  \right)  >0$; When $\rho_{\gamma}$ can be
ignored,
\begin{equation}
k\left(  \rho\right)  =1,\text{ \ for }\rho_{m1}>0;\text{ \ }k\left(
\rho\right)  =-1,\text{ \ for }\rho_{m1}<0. \tag{D4d}%
\end{equation}

Because $k\left(  \rho\right)  =k\left(  \rho\left(  t,R(t)\right)  \right)
$, the corresponding equations should be deduced from $(D2)$. As a rough
approximation, we have in the international unit%
\begin{equation}
\overset{\cdot}{R}^{2}=-k\left(  \rho\right)  c^{2}+\eta\left(  \rho_{m}%
+\rho_{\gamma}\right)  R^{2}, \tag{D5a}%
\end{equation}%
\begin{equation}
\overset{\cdot\cdot}{R}=-\frac{c^{2}dk}{2\overset{\cdot}{R}dt}-\frac{1}{2}%
\eta\left(  \rho_{m}+2\rho_{\gamma}\right)  R=-\frac{c^{2}}{2}\left(
\frac{\partial k}{\overset{\cdot}{R}\partial t}+\frac{\partial k}{\partial
R}\right)  -\frac{1}{2}\eta\left(  \rho_{m}+2\rho_{\gamma}\right)  R,
\tag{D5b}%
\end{equation}
where $p_{\gamma}=\rho_{\gamma}/3$ is considered. When $V=\partial\rho
_{m}/\partial t=\partial\rho_{\gamma}/\partial t=0,$ from $(D5)$ we can still
obtained $\rho_{m}R^{3}=\rho_{m1}R_{1}^{3}$ and $\rho_{\gamma}R^{4}%
=\rho_{\gamma1}R_{4}^{4}.$ In the case, let $H_{1}^{2}=\left(  \overset{\cdot
}{R}_{1}/R_{1}\right)  ^{2}=\eta\rho_{c1},$ $\Omega_{1}=\rho_{1}/\rho_{c1}$
for $\rho_{1}>0$ and $\Omega_{1}=-\rho_{1}/\rho_{c1}$ for $\rho_{1}<0,$
$\Omega_{m1}=\rho_{m1}/\rho_{c1}$ for $\rho_{m1}>0$ and $\Omega_{m1}%
=-\rho_{m1}/\rho_{c1}$ for $\rho_{m1}<0,$ and $a=R/R_{1},$ we can simplify
$(D5)$ to%
\begin{align}
\overset{\cdot}{a}^{2}  &  =\frac{\Omega_{m1}-\Omega_{\gamma1}/a}{R_{1}%
^{2}\Omega_{1}}c^{2}-H_{1}^{2}\left(  \frac{\Omega_{m1}}{a}-\frac
{\Omega_{\gamma1}}{a^{2}}\right)  =\frac{c^{2}}{R_{1}^{2}}\widetilde{k}%
-H_{1}^{2}\left(  \frac{\widetilde{\Omega}_{m1}}{a}-\frac{\Omega_{\gamma1}%
}{a^{2}}\right) \nonumber\\
&  =\frac{\Omega_{m1}c^{2}}{R_{1}^{2}\Omega_{1}}\left(  1-\frac{P_{1}}%
{a}\right)  \left(  1-\frac{Q_{1}}{a}\right)  , \tag{D6a}%
\end{align}%
\begin{equation}
\overset{\cdot\cdot}{a}=\frac{H_{1}^{2}}{2}\left[  \frac{\widetilde{\Omega
}_{m1}}{a^{2}}-2\frac{\Omega_{\gamma1}}{a^{3}}\right]  =\frac{H_{1}^{2}%
\Omega_{m1}}{2Q_{1}a^{2}}\left[  P_{1}\left(  1-\frac{Q_{1}}{a}\right)
+Q_{1}\left(  1-\frac{P_{1}}{a}\right)  \right]  , \tag{D6b}%
\end{equation}
where
\begin{align}
P_{1}  &  =\Omega_{\gamma1}/\Omega_{m1},\text{ \ \ }\widetilde{k}=\Omega
_{m1}/\Omega_{1},\text{\ \ }\widetilde{\Omega}_{m1}\equiv\Omega_{m1}\left(
1+\frac{P_{1}}{Q_{1}}\right)  .\tag{D7a}\\
Q_{1}  &  =H_{1}^{2}R_{1}^{2}\Omega_{1}/c^{2}=\Omega_{1}/\left(  (1+\Omega
_{1}\right)  ,\text{ \ for \ }\rho_{1}<0,\tag{D7b}\\
&  =\Omega_{1}/\left(  (\Omega_{1}-1\right)  ,\text{ \ \ \ for \ }\rho_{1}>1.
\tag{D7c}%
\end{align}
We can arbitrarily choose $t_{1}$ provided $\rho\left(  t_{1}\right)
=\rho_{v}\left(  t_{1}\right)  -\rho_{s}\left(  t_{1}\right)  =\rho_{1}\neq0.$

In order for convenience, we may take
\begin{align}
k  &  =1\text{ \ \ for \ \ }\rho>0,\tag{D8a}\\
k  &  =0\text{ \ \ for \ \ }\rho=0,\tag{D8b}\\
k  &  =-1\text{ \ for \ \ }\rho<0. \tag{D8c}%
\end{align}

From $(D5)$ we can obtain the results similar to $(84)$, $(86)$ and figures 1
and 2.

\textbf{Acknowledgement}

I am very grateful to professor Zhao Zhan-yue for his helpful discussions and
best support.


\begin{thebibliography}{99}                                                                                               %


\bibitem {[1]}S. W. Hawking and G.F.R. Ellis, The Large Scale Strcture of
Space-Time,Cambidge University Press, 1999, p256-298.

\bibitem {[2]}R.R.Caldwell, Phys. World \textbf{17}, 37 (2004); T.padmanabhan,
Phys. Rep. \textbf{380}. 325 (2003); P.J.E. Peebles and B. Ratra; Rev. Mod.
Phys. \textbf{75},559 (2003).

\bibitem {[3]}S. Weinberg, Phys. Rev. Lett. \textbf{59}, 2607 (1987);
H.Martel, P.R. Shapiro and S. Weinberg, Astrophys.J \textbf{492}, 29 (1998).

\bibitem {[4]}P.J.E. Peebles and B. Ratra, Astrophys. \textbf{J}.\textbf{325},
L17 (1988); B. Ratra and P.J.E. Peebles, Phys. Rev. D \textbf{37}, 3406
(1988); P.J.E. Peebles and B.Ratra, Rev. Mod. Phys. \textbf{75}. 559 (2003).
N.Weiss, Phys. Lett. \textbf{B 197}, 42 (1987).

\bibitem {[5]}L.J. Hall, Y. Nomura and S.J. Oliver, Phys.. Rev. Lett.
\textbf{95}, 141302 (2005).

\bibitem {[6]}S-H. Chen, `Quantum Field Theory :New Research', O. Kovras
Editor, Nova Science Publishers, Inc. (2005), p103-170.

\bibitem {[7]}M.Chaichian and N.F.Nelipa `Introduction to Gauge Field
Theories' Springer-Verlag Berlin Heidelberg, 1984, p269; Graham G.Ross, `Grand
Unified Theories', The Benjamin/Cummings Publishing Company, INC, 1984, p177-183.

\bibitem {[8]}S. Weinberg, Gravitation and Cosmology, chanter 12 section 3,
New York, Wiley, 1972.

\bibitem {[9]}Liu Liao, Jiang Yuanfang and Qian Zhenhua, Progress in Physics,
V.9, No. 2, (1989), p159.

\bibitem {[10]}S.Coleman and E.J.Weiberg Phys. Rev.D7. 1888 (1973).
R.H.\ Brandenberg, Rev. of Mod. Phys. 57. 1 (1985).

\bibitem {[11]}A. H. Guth, Phys. Rev. D 23, 347 (1981).

\bibitem {[12]}J. A. Peacock, Cosmological Physics, Cambridge University Press
(1999), (3.44) in p78, (3.81) in p90, (3.78) in p89, p664.

\bibitem {[13]}H. C. Ohanian and R. Ruffini, Gravitation and Spacetime (2nd
ed.), W.W. Norton and Company, Inc. (1994), Section 9.7.
\end{thebibliography}
\end{document}